\documentclass[aps,prc,twocolumn,showpacs,floatfix,nofootinbib,preprintnumbers,superscriptaddress,longbibliography,amsproc,amsmath,amssymb]{revtex4-2}

\usepackage{epsfig,dsfont,amssymb,amsmath,amsthm,amsfonts,amsbsy,mathrsfs}
\usepackage{graphicx}

\usepackage{amsmath}
\usepackage{amssymb}%
\usepackage{dcolumn}
\usepackage{hyperref}
\usepackage{soul} 
\graphicspath{{.}}

\usepackage[disable]{todonotes}

\newcommand{\be}{\begin{equation}}
\newcommand{\ee}{\end{equation}}
\newcommand{\ba}{\begin{eqnarray}}
\newcommand{\ea}{\end{eqnarray}}

\begin{document}

\title{Two-pion exchange as a leading-order contribution in chiral effective field theory}

\author{Chinmay Mishra}
\affiliation{Department of Physics and Astronomy, University of
  Tennessee, Knoxville, Tennessee 37996, USA}

\author{A.~Ekstr{\"o}m}

\affiliation{Department of Physics, Chalmers University of Technology,
  SE-412 96 G{\"o}teborg, Sweden}

\author{G.~Hagen}

\affiliation{Physics Division, Oak Ridge National Laboratory, Oak
  Ridge, Tennessee 37831, USA}

\affiliation{Department of Physics and Astronomy, University of
  Tennessee, Knoxville, Tennessee 37996, USA}

\author{T.~Papenbrock}

\affiliation{Department of Physics and Astronomy, University of
  Tennessee, Knoxville, Tennessee 37996, USA}

\affiliation{Physics Division, Oak Ridge National Laboratory, Oak
  Ridge, Tennessee 37831, USA}

\author{L.~Platter}

\affiliation{Department of Physics and Astronomy, University of
  Tennessee, Knoxville, Tennessee 37996, USA}

\affiliation{Physics Division, Oak Ridge National Laboratory, Oak
  Ridge, Tennessee 37831, USA}

\affiliation{Institut f\"ur Kernphysik, Technische Universit\"at Darmstadt, 64289 Darmstadt, Germany}

\begin{abstract}
  Pion exchange is the central ingredient to nucleon-nucleon
  interactions used in nuclear structure calculations, and one pion exchange (OPE) enters at
  leading order in chiral effective field theory.  In the $^{2S+1}L_J={^1S_0}$ partial wave, however,
  OPE and a contact term needed for proper renormalization fail to
  produce the qualitative, and quantitative, features of the scattering phase shifts. Cutoff variation also revealed a surprisingly low breakdown momentum $\Lambda_\text{b}\approx 330$~MeV in this partial wave. Here we show that potentials consisting of OPE, two pion exchange (TPE), and a single contact address these problems and yield accurate and renormalization
  group (RG) invariant phase shifts in the $^1S_0$ partial
  wave. We demonstrate that a leading-order potential with TPE can be systematically improved by adding a contact quadratic in momenta. 
  For momentum cutoffs $\Lambda \lesssim 500$~MeV, the removal of relevant physics from TPE loops needs to be
  compensated by additional contacts to keep RG invariance. Inclusion of the $\Delta$ isobar degree of freedom in the potential does not change the strong contributions of TPE.
\end{abstract}


\maketitle

\section{Introduction}
Ever since its introduction by \textcite{yukawa1935}, boson exchange has been central to the theory of nuclear interactions. In quantum chromodynamics, chiral symmetry is spontaneously and explicitly broken, and the pion emerges as the corresponding pseudo Nambu-Goldstone boson. Pion exchange, together with contact interactions that account for unknown short-range physics, thus are the ingredients in a chiral effective field theory
(EFT) description of the nucleon-nucleon interaction~\cite{weinberg1990,weinberg1991,ordonez1992,vankolck1994,kaiser1997,epelbaum2000,epelbaum2009,Hammer:2019poc}.

In chiral EFT, and within the commonly employed Weinberg power counting, the leading-order contributions to the nucleon-nucleon interaction consist of OPE and one contact each in the $^1S_0$ and $^3S_1$ partial waves. At next-to-leading order (NLO) in Weinberg counting the leading TPE contributions as well as contacts quadratic in momenta enter.

Statistical analyses of higher-order chiral EFT predictions for nucleon-nucleon scattering data infer a breakdown momentum $\Lambda_\text{b}\approx 600-700$~MeV, and that higher chiral orders yield systematical improvements in powers of $Q/{\rm min}(\Lambda,\Lambda_{\rm b})$ beyond the leading-order results~\cite{reinert2018,wesolowski2019}. Here $Q$ is the low-momentum scale of interest, e.g., the external momentum in nucleon-nucleon scattering, and $\Lambda$ is the cutoff employed in the regularization of the theory. While this looks encouraging, there are well-known challenges~\cite{furnstahl2021}, and we mention two of them.

First, recent results~\cite{carlsson2016,yang2021,maris2021} show that leading-order chiral EFT potentials predict light-mass nuclei that are unstable with respect to breakup into $\alpha$ particles and lighter-mass clusters, raising questions about what should be expected from a nuclear EFT at leading order. Of course, the lack of any spin-orbit contributions at leading order in the Weinberg power counting would also presumably make well-known nuclear shell-structure a subleading effect.

\begin{figure}[htb]
  \includegraphics[width=0.5\textwidth]{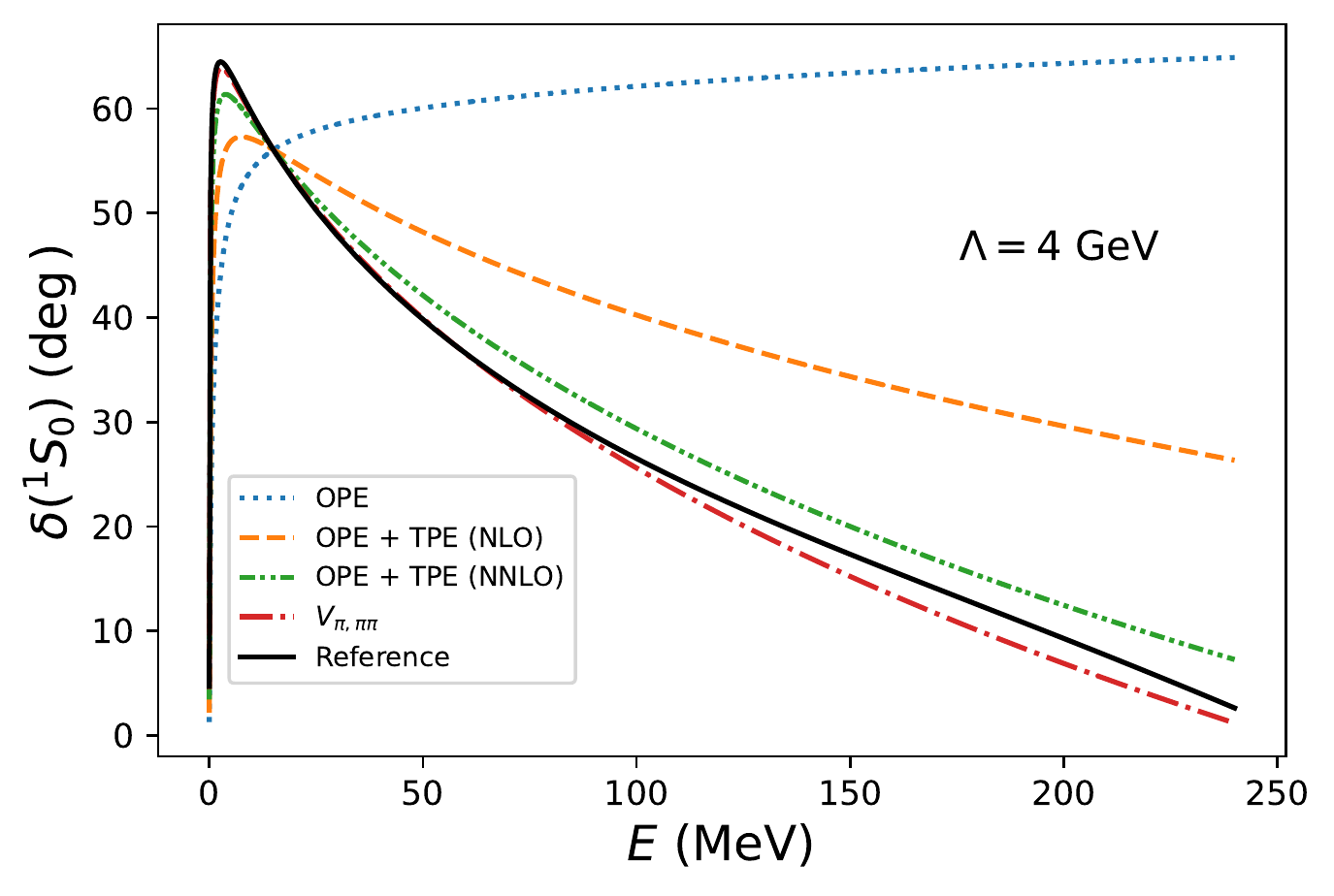}
  \caption{(Color online) Nucleon-nucleon phase shifts in the $^1S_0$ partial wave versus the laboratory scattering energy $E$ for various chiral interactions. The momentum cutoff is set to $4$~GeV and in all cases a single contact $V_{\rm ct}^{(0)}$ is adjusted to match the  phase shift to that of the reference potential (high-precision Idaho-N3LO) by~\textcite{entem2003} (solid black line) at $E_0=15$~MeV. The dotted blue line shows phase shift obtained for OPE, the dashed orange line for OPE plus the leading TPE, the dash-dot-dotted green line for OPE plus subleading TPE, and the dash-dotted red line for the interaction $V_{\pi,\pi\pi}$ consisting of OPE and leading plus subleading TPE (see text for details). 
  \label{fig:compare-chiral-pots}}
\end{figure}

Second, the leading-order description of nucleon-nucleon phase shifts in the $^1S_0$ partial wave are problematic in the Weinberg power
counting~\cite{kaplan1998,frederico1999,nogga2005,valderrama2005,valderrama2006,birse2006,shukla2008,yang2008,birse2010,long2012,long2012b,valderrama2015,sanchez2018,odell2019,sanchez2020,yang2021}, see
Ref.~\cite{vankolck2020} for a recent review. The combination of OPE and a single contact fails qualitatively to capture the pronounced peak at $60-70$ degrees in the phase shift at about 8~MeV of laboratory scattering energy, see blue dotted line in Fig.~\ref{fig:compare-chiral-pots}. For all results in this figure, the $^1S_0$ contact is adjusted to the reference phase shifts of the potential~\cite{entem2003} (shown as a solid black line) at a laboratory scattering energy $E=15$~MeV. This energy matches the relevant scale of pion physics as $m_\pi^2/m_N\approx 20$~MeV using the pion and nucleon mass $m_\pi$ and $m_N$, respectively. The OPE phase shifts are much too attractive beyond the matching point and clearly fail to capture the characteristics of the reference phase shifts. Also, the slope of the OPE phase shift has the wrong sign.

Another problem concerns the breakdown momentum $\Lambda_\text{b}$. The analysis of the $^1S_0$ phase shifts by~\textcite{lepage1997} showed that a potential consisting of OPE plus leading and subleading contacts leads to $\Lambda_\text{b}\approx 330$~MeV in the $^1S_0$ partial wave. Later analyses exploring perturbative inclusion of subleading TPE contributions~\cite{birse2006,valderrama2011} confirmed this finding and estimated the breakdown momentum to be even lower, i.e., $\Lambda_{\rm b}\approx 200$~MeV. This questions whether leading-order chiral EFT based on OPE physics is consistent with the general assumption that the breakdown momentum is somewhere between $\sim 500$ and $\sim 1000$~MeV.

Several researchers addressed the shortcoming of too attractive $^1S_0$ phase shifts by adding effective-range corrections~\cite{epelbaum2015b,yang2021}, energy-dependent potentials generated by di-baryon fields~\cite{sanchez2018}, or separable potentials~\cite{sanchez2020} to the OPE as leading-order contributions. These approaches improve the phase shifts at a cost of introducing additional parameters. The promotion of TPE to leading order in chiral EFT was proposed already a decade ago by~\textcite{birse2010}, however, a rigorous analysis has (to the best of our knowledge) never been carried out. 

In this paper we show that chiral physics in the form of TPE, presumably a higher-order correction in the Weinberg power counting, remedies the shortcomings of highly attractive $^1S_0$ phase shifts and without increasing the number of parameters. As Fig.~\ref{fig:compare-chiral-pots} shows, the inclusion of TPE at leading order significantly improves the accuracy of the $^1S_0$ phase shifts. We will also show that the estimated breakdown momentum in this approach is consistent with expectations from chiral EFT. 

Phenomenology connects TPE with the strong mid-range attraction of the nucleon-nucleon force~\cite{wiringa1995,machleidt2001} which is also attributed to the $f_0(500)$ resonance~\cite{pelaez2016} or the sigma meson whose width and mass was determined model-independently in Ref.~\cite{Caprini:2005zr}. The latter is a central ingredient of relativistic mean-field theories~\cite{walecka1986,reinhard1989,gambhir1990,serot1992,sugahara1994} and of 
alternative proposals to chiral EFT which include the effects of the $f_0(500)$ resonance at leading order via the sigma meson~\cite{donoghue2006} or the dilaton~\cite{crewther2015,ma2020}, i.e. the Nambu-Goldstone boson of a broken and hidden scale symmetry.

In the Weinberg power counting TPE enters at NLO, but its strongest contributions involving the pion-nucleon coupling constants $c_i$, with $i=1,3,4$, enter at next-to-next-to-leading order (NNLO). It is known that these subleading TPE contributions are crucial for quantitatively reproducing the $^1S_0$ phase shifts, see, e.g., Refs.~\cite{valderrama2006,machleidt2011}. We note, however, that the role of TPE is somewhat
obscured in chiral potentials. First, an additional contact potential quadratic in momentum also enters in the $^1S_0$ partial wave at NLO in the Weinberg power counting. Second, several popular
potentials~\cite{entem2003,ekstrom2013,ekstrom2015a,entem2015,jiang2020} employ relatively low momentum cutoffs and thereby truncate some parts of the TPE strength.

This manuscript is ordered as follows: In Sec.~\ref{sec:ope-tpe-potentials} we give the
explicit form of the OPE and TPE potentials, discuss their relative strengths, and show results for  phase shifts in the $^1S_0$ partial wave obtained using the proposed leading-order potential where we have promoted TPE. In Sec.~\ref{sec:subleading} we show  that a subleading contact that is quadratic in momenta systematically improves upon these results. In Sec.~\ref{sec:delta} we show that the inclusion of $\Delta$ isobar degrees of freedom does not alter our conclusions about promoting TPE to leading order. We end with a summary in Sec.~\ref{sec:summary}.

\section{Chiral OPE and TPE potentials}
\label{sec:ope-tpe-potentials}
Canonical chiral EFT descriptions of the nuclear interaction employ a power counting for the pion-nucleon potential as done in chiral perturbation theory. The OPE enters at leading order, and subleading contributions are presumably suppressed by powers of $g_A m_\pi/(4\pi f_\pi)\ll 1$, where $g_A \approx 1.28$ is the axial-vector
constant, $m_\pi\approx 140$~MeV is the pion mass, and $f_\pi\approx 92$~MeV is the pion-decay constant. The OPE potential is given by
\begin{equation}
  \label{ope}
  V_{\rm OPE}(\mathbf{q})= -{g_A^2\over 4f_\pi^2} \boldsymbol{\tau}_1\cdot \boldsymbol{\tau}_2
  \frac{(\boldsymbol{\sigma}_1\cdot\mathbf{q}) (\boldsymbol{\sigma}_2\cdot\mathbf{q})}{m_\pi^2+q^2} \ .
\end{equation}
Here, $\mathbf{q}=\mathbf{p}'-\mathbf{p}$ is the momentum transfer. The Pauli matrices of nucleon $j$ in spin and isospin space are denoted as $\mathbf{\sigma}_j$ and $\mathbf{\tau}_j$, respectively. The TPE potentials considered in this work can be written as 
\begin{eqnarray}
  \label{tpe}
  V_{\rm TPE}(\mathbf{q}) &=& V_C(q) + \boldsymbol{\tau}_1\cdot \boldsymbol{\tau}_2 W_C(q) \nonumber\\
  &+&\left[V_T(q) + \boldsymbol{\tau}_1\cdot \boldsymbol{\tau}_2 W_T(q)\right] (\boldsymbol{\sigma}_1\cdot\mathbf{q}) (\boldsymbol{\sigma}_2\cdot\mathbf{q}) \nonumber\\
  &+&\left[V_S(q) + \boldsymbol{\tau}_1\cdot \boldsymbol{\tau}_2 W_S(q)\right] \boldsymbol{\sigma}_1\cdot\boldsymbol{\sigma}_2 \ . 
\end{eqnarray}
The leading TPE potential, i.e., the contributions that enter at NLO in the Weinberg power counting is given by~\cite{kaiser1997,epelbaum2000,machleidt2011}
\begin{equation}
  \label{nlo}
    \begin{aligned}
  W_C&=-{L(q)\over 384\pi^2 f_\pi^4} \bigg[4m_\pi^2(5g_A^4-4g_A^2-1)\\
    & + q^2(23g_A^4-10g_A^2-1)+{48g_A^4m_\pi^4\over w^2}\bigg] \ , \\
  V_T&=-{V_S\over q^2}=-{3g_A^4 L(q)\over 64\pi^2 f_\pi^4} \ , 
    \end{aligned}
\end{equation}
while the subleading contributions that enter at NNLO are~\cite{kaiser1997,epelbaum2000,machleidt2011}
\begin{equation}
  \label{nnlo}
  \begin{aligned}
  V_C&=-{3g_A^2\over 16\pi f_\pi^4} \left[2m_\pi^2(2c_1-c_3)-c_3q^2\right]\tilde{w}^2A(q) \ , \\ 
  W_T&=-{W_S\over q^2}=-{g_A^2\over 32\pi f_\pi^4} c_4 w^2 A(q)\ .
  \end{aligned}
\end{equation}
Here, we used the short-hands
\begin{align}
  w&\equiv \sqrt{4m_\pi^2+q^2} \ , \\
  L(q)&\equiv {w\over q}\log{w+q\over 2m_\pi} \ , \\
  \tilde{w}&\equiv \sqrt{2m_\pi^2+q^2} \ , \\
  A(q)&\equiv {1\over 2q}{\rm arctan}{q\over 2m_\pi} \ ,
\end{align}
with $q\equiv |\mathbf{q}|$, and the pion-nucleon constants $c_i$ are of the order of $m_N^{-1}$ with $m_N$ denoting the nucleon mass $m_N$. In what follows we use $c_1=-0.74$~GeV$^{-1}$, $c_3=-3.61$~GeV$^{-1}$, and $c_4=2.44$~GeV$^{-1}$ from Ref.~\cite{reinert2018}, obtained from Roy-Steiner relations~\cite{hoferichter2015,hoferichter2016}. We note here that we have neglected any relativistic corrections (proportional to $m_N^{-1}$) to the TPE at NNLO, and we have omitted any polynomial contributions.

\begin{figure}[htb]
  \includegraphics[width=0.5\textwidth]{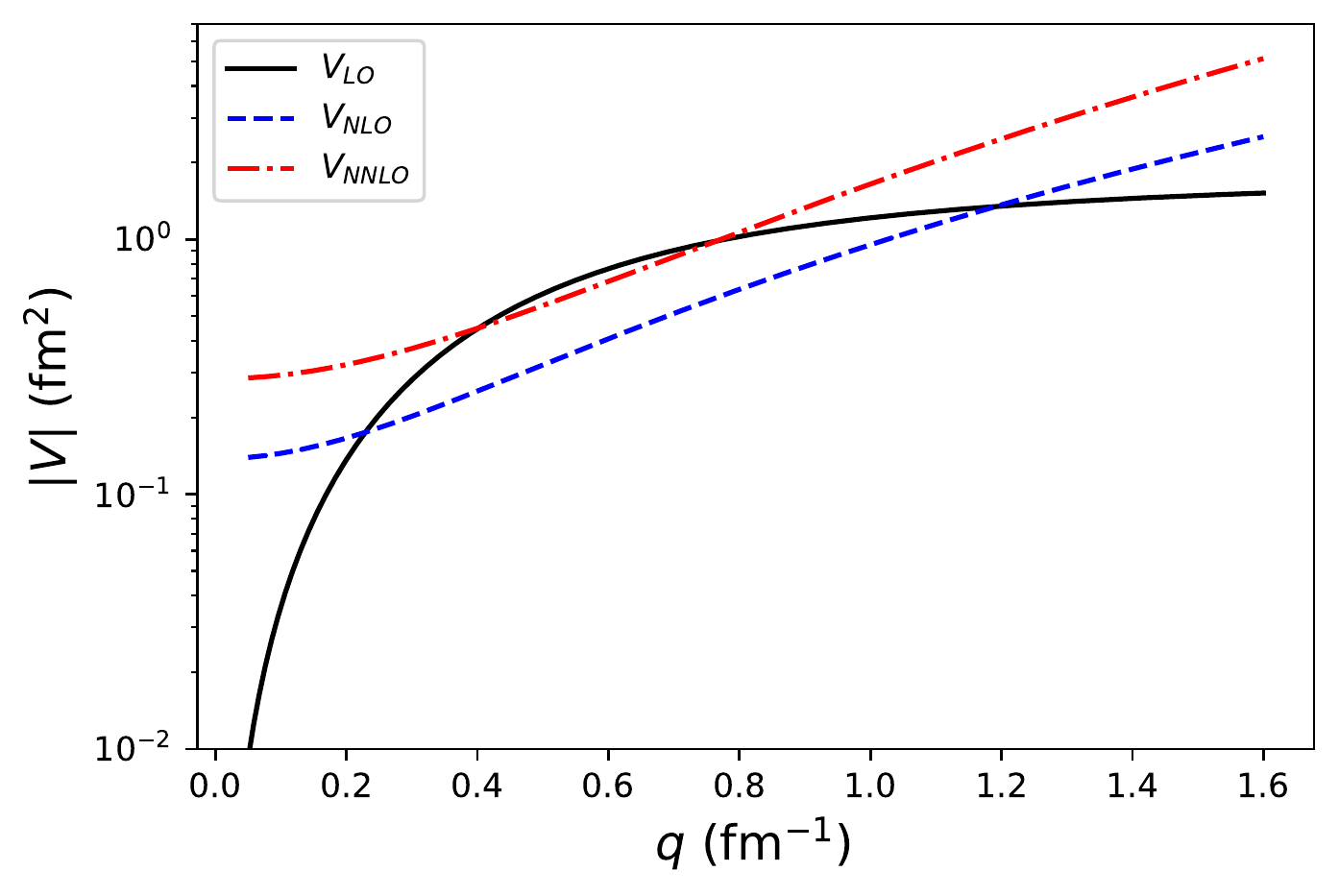}
  \caption{(Color online) Pion-exchange contributions to the
    nucleon-nucleon potentials in the spin-singlet / isospin-triplet partial wave at various  orders in Weinberg power counting as a function of momentum transfer. The OPE potential (black solid line) is leading order in the Weinberg power counting, the leading TPE (dashed blue line) is NLO, and the subleading TPE (dash-dotted red line) enters at NNLO.
  \label{fig1}}
\end{figure}

We evaluated the potentials in the spin-singlet / isospin-triplet partial wave and show their magnitudes in Fig.~\ref{fig1} comparing the OPE, leading TPE, and subleading TPE contributions.    
As expected, OPE is a dominant contribution, while the subleading TPE potential of Eq.~(\ref{nnlo}) cannot be neglected around momentum transfers of about 1~fm$^{-1}$. Clearly, at momentum transfers of the order of the pion mass ($m_\pi\approx 0.7$~fm$^{-1}$) or the Fermi momentum in nuclear matter at saturation ($k_F\approx 1.35$~fm$^{-1}$), the placement of this TPE potential at NNLO does not reflect its actual strength.

Let us study the impact of TPE added to OPE in the $^1S_0$ phase shifts. We define a nucleon-nucleon potential in this partial wave that is of the form
\begin{equation}
    V(\mathbf{p}',\mathbf{p}) \equiv V_{\pi,\pi\pi} + V_{\rm{ct}} \ .
\end{equation} 
Here, $V_{\pi,\pi\pi}$ consists of OPE and the leading and subleading TPE, while $V_{\rm ct}$ denotes the contact potential to be specified. At leading order and NLO, the contact potentials are given by
\begin{equation}
V_{\rm ct}^{(0)} (\mathbf{p'},\mathbf{p}) = \tilde{C} \ , 
\label{contact0}
\end{equation}
and 
\begin{equation}
V_{\rm ct}^{(2)}  (\mathbf{p'},\mathbf{p}) = C (p'^2 + p^2) \ ,    
\label{contact2}
\end{equation}
respectively. The potentials are regularized via
\begin{align}
V \to f({p^\prime}^2/\Lambda^2) V f(p^2/\Lambda^2)
\end{align}
using $p\equiv|\mathbf{p}|$, $p'\equiv |\mathbf{p}'|$, 
\begin{equation}
\label{regulator}
f(x)=e^{-x^n} \ ,
\end{equation}
and $n=3$. For TPE loop diagrams we use a spectral function regulator with a cutoff of
700~MeV as introduced in~\cite{Epelbaum:2003gr,epelbaum04}. 

At leading order we adjust the low-energy constant (LEC) $\tilde{C}$ such that the phase shift at the laboratory energy $E_0=15$~MeV reproduces the value from the high-precision Idaho-N3LO potential~\cite{entem2003}, which we take as a reference throughout this work. The precision of this reference is sufficient for our purposes, as the phase shifts of this potential are virtually indistinguishable from a recent partial-wave analysis of nucleon-nucleon scattering data~\cite{navarro2013}. 

The $^1S_0$ phase shifts for our leading-order potential are shown as the red dash-dotted line in Fig.~\ref{fig:compare-chiral-pots}. For comparison, we also show the results from potentials with other combinations of pion exchanges; such as the sum of OPE and leading TPE (orange dashed line) and OPE plus subleading TPE (green dash-dot-dotted line). The chiral potential $V_{\pi, \pi \pi}$ stands out through its accuracy for scattering energies below and above the energy $E_0$ used for renormalizing the contact LEC. This suggests that the combination $V_{\pi,\pi\pi}$ of OPE and TPE should be taken as the leading-order contribution from chiral physics in the $^1S_0$ partial wave. This is the main result of this paper. It is consistent with the anticipation obtained from Fig.~\ref{fig1}. Although the combination of OPE and leading TPE (shown as the orange dashed line) brings the phase shift closer to reference it fails to reproduce the characteristic decrease of the phase shifts with increasing energy, i.e., a lack of increasing repulsion with $E$, and the amplitude zero is nowhere near $E\approx 250$~MeV. In contrast, using $V_{\pi,\pi\pi}$ leads to a leading-order nucleon-nucleon potential that captures the main qualitative and quantitative features of the $^1S_0$ phase shifts.

\begin{figure}[htb]
\centering
\includegraphics[width=0.5\textwidth]{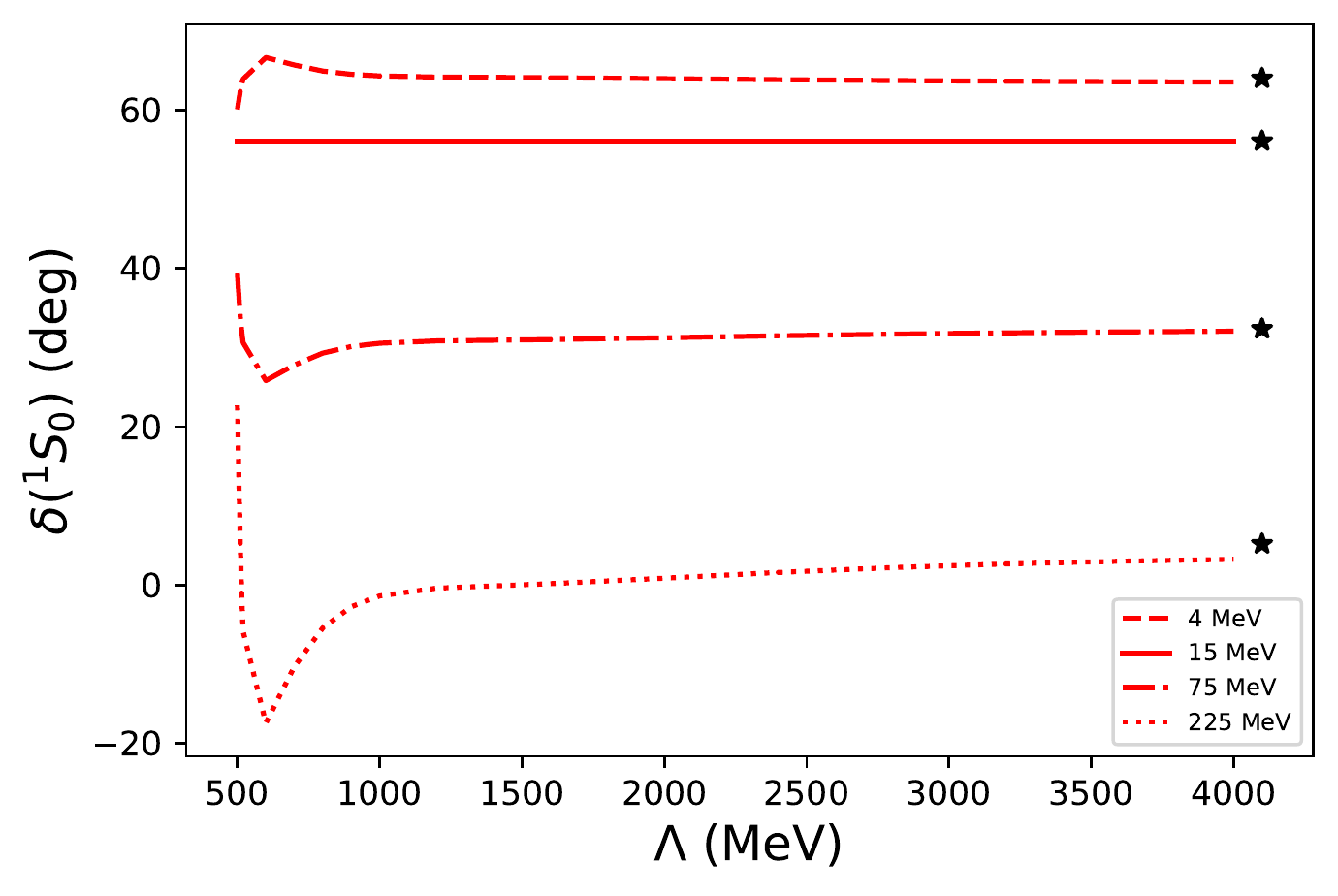}
\caption{(Color online) $^1S_0$ phase shifts as a function of the momentum regulator cutoff $\Lambda$ at laboratory energies of $E=4$ (dashed), $15$  (solid), $75$ (dash-dotted), and $225$~MeV (dotted) for the potential $V_{\pi,\pi\pi}+V_{\rm ct}^{(0)}$. The black stars show the reference phase shifts at these energies.}
\label{fig:RGI}
\end{figure}

\begin{figure}[!htbp]
    \includegraphics[width=\linewidth]{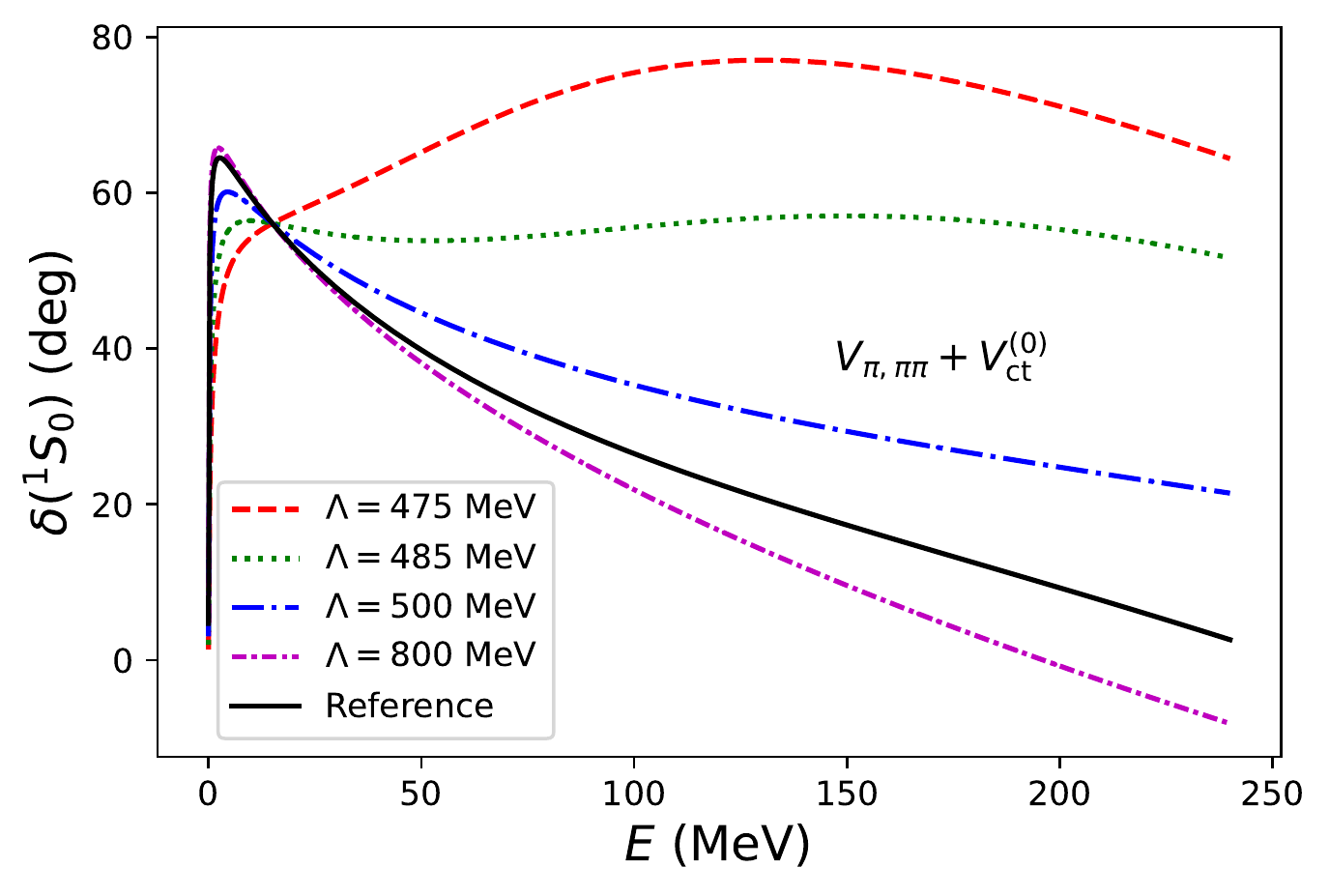}
    \caption{(Color online) Phase shifts as a function of laboratory energy of the interaction $V_{\pi,\pi\pi}+V_{\rm ct}^{(0)}$ at low momentum cutoffs as indicated and compared to the reference (solid black line). The phase shifts become increasing repulsive as the cutoff is increased. For $E>15$~MeV they cross over and become more repulsive than the reference for cutoffs in the range $500 \lesssim \Lambda \lesssim 520$~MeV.}
    \label{fig:small}
\end{figure}

We note that the phase shifts presented in Fig.~\ref{fig:compare-chiral-pots} include chiral physics plus a single LEC that accounts for unknown short-range physics. Figure~\ref{fig:RGI} demonstrates that the potential $V_{\pi,\pi\pi} + V_{\rm ct}^{(0)}$ yields RG invariant $^1S_0$ phase shifts as the cutoff $\Lambda$ is increased. For a comparison, the reference phase shifts are shown as black stars. 
Figure~\ref{fig:RGI} also shows that the phase shifts are strongly cutoff dependent for $\Lambda \lesssim 750$ MeV. This can be understood as follows: In coordinate space, the OPE potential exhibits a $1/r^3$ behavior at short distances, where $r$ denotes the two-nucleon relative distance. The TPE potential exhibits a significantly more singular $1/r^6$ short-distance behavior with a typical momentum scale $k_{\pi\pi}\sim 115$~MeV set by the chiral couplings employed in the TPE potential~\cite{valderrama2011}. The cutoff needs therefore to be significantly larger than $k_{\pi\pi}$ to reach convergence. Alternatively, the relevant scale of the TPE interaction can also naively be estimated by 
\begin{equation}
  \label{lam_tpe}
\Lambda_\text{TPE}\equiv \sqrt{2m_\pi
  m_N}\approx 510~\mbox{MeV} \ .
\end{equation}
Cutoffs lower than $\Lambda_{\rm TPE}$ therefore remove physics from the TPE. This effect is highlighted in Fig.~\ref{fig:small}. As before, we adjusted the leading-order contact $V_{\rm ct}^{(0)}$ to reproduce the reference phase shift at $E_0=15$~MeV. We clearly see the effects of removing TPE physics for cutoffs $\Lambda \lesssim 500$~MeV. Indeed, comparison with Fig.~\ref{fig:compare-chiral-pots} shows that the phase shifts at a cutoff of $\Lambda=475$~MeV are close to those of OPE plus a contact. As we will see below, adding the subleading contact $V_{\rm ct}^{(2)}$ will restore RG invariance in this case.     

Figure~\ref{fig:small} also allows us to estimate the breakdown momentum in this partial wave as the momentum regulator cutoff value for which the phase shift predictions are closest to the reference~\cite{lepage1997}. Following this strategy we infer $\Lambda_\text{b}\approx 500-520$~MeV. This is significantly larger than what was found when only OPE physics is included at leading order~\cite{lepage1997,birse2006,valderrama2011}. It is also in line with expectations from neglecting the physics of more massive exchange mesons.

\section{Systematic improvements}
\label{sec:subleading}

We propose to employ higher-order contacts to systematically improve upon the results of our leading-order potential $V_{\pi,\pi\pi}+V_{\rm ct}^{(0)}$. Thus, the contact~(\ref{contact2}) enters as the subleading correction~\cite{lepage1997}. This introduces a new LEC, $C$, and we adjust $\tilde{C}$ and $C$ such that 
the reference phase shift and its slope is reproduced at the laboratory energy $E_0=15$~MeV. In our numerical work, we used the secant slope between $E_0$ and a second point just below this energy rather than the exact tangent slope. We work at a cutoff of $\Lambda=800$~MeV.

The question then arises whether one should treat the subleading correction perturbatively, or not. In the non-perturbative approach, the full potential is iterated to solve the Lippmann Schwinger equation, while the perturbative approach is linear in the subleading correction. We followed both approaches and found similar results for the phase shifts. 

\begin{figure}[!htbp]
    \includegraphics[width=\linewidth]{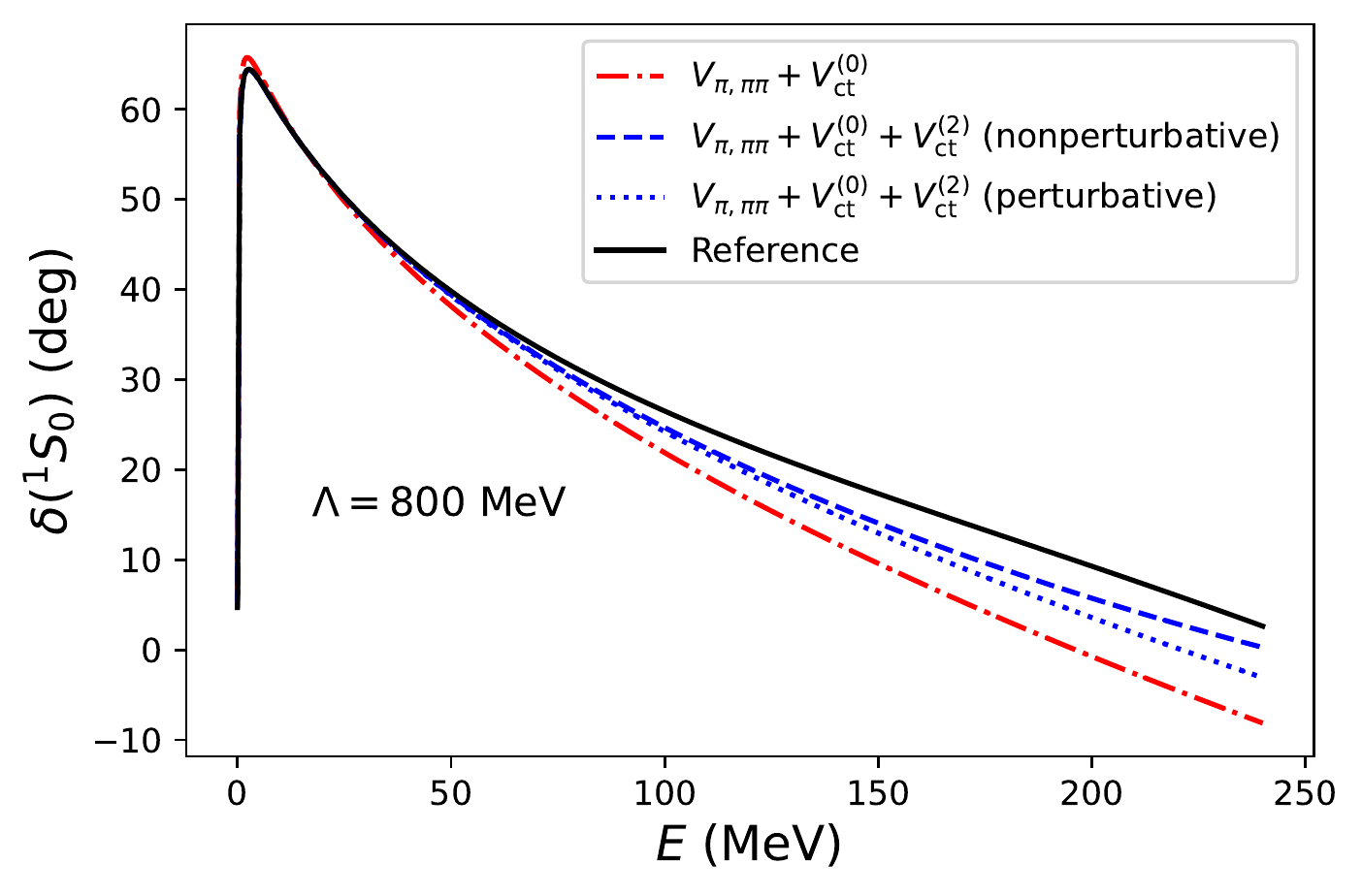}
    \caption{(Color online) Systematic improvements of the $^{1}S_0$ phase shifts by adding the subleading contact $V_{\rm ct}^{(2)}$ to the  potential $V_{\pi,\pi\pi}+V_{\rm ct}^{(0)}$. The cutoff is $800$~MeV. Results for the potentials $V_{\pi,\pi\pi}+V_{\rm ct}^{(0)}$ and $V_{\pi,\pi\pi}+V_{\rm ct}^{(0)}+V_{\rm ct}^{(2)}$ are shown as a dash-dotted red line and as blue lines, respectively. The dashed and the dotted blue line correspond to the  non-perturbative and the perturbative inclusion of the subleading contact $V_{\rm ct}^{(2)}$, respectively, and the solid black line shows the reference.}
  \label{fig:LOvNLO}
\end{figure}

\begin{figure}[htb]
    \includegraphics[width=\linewidth]{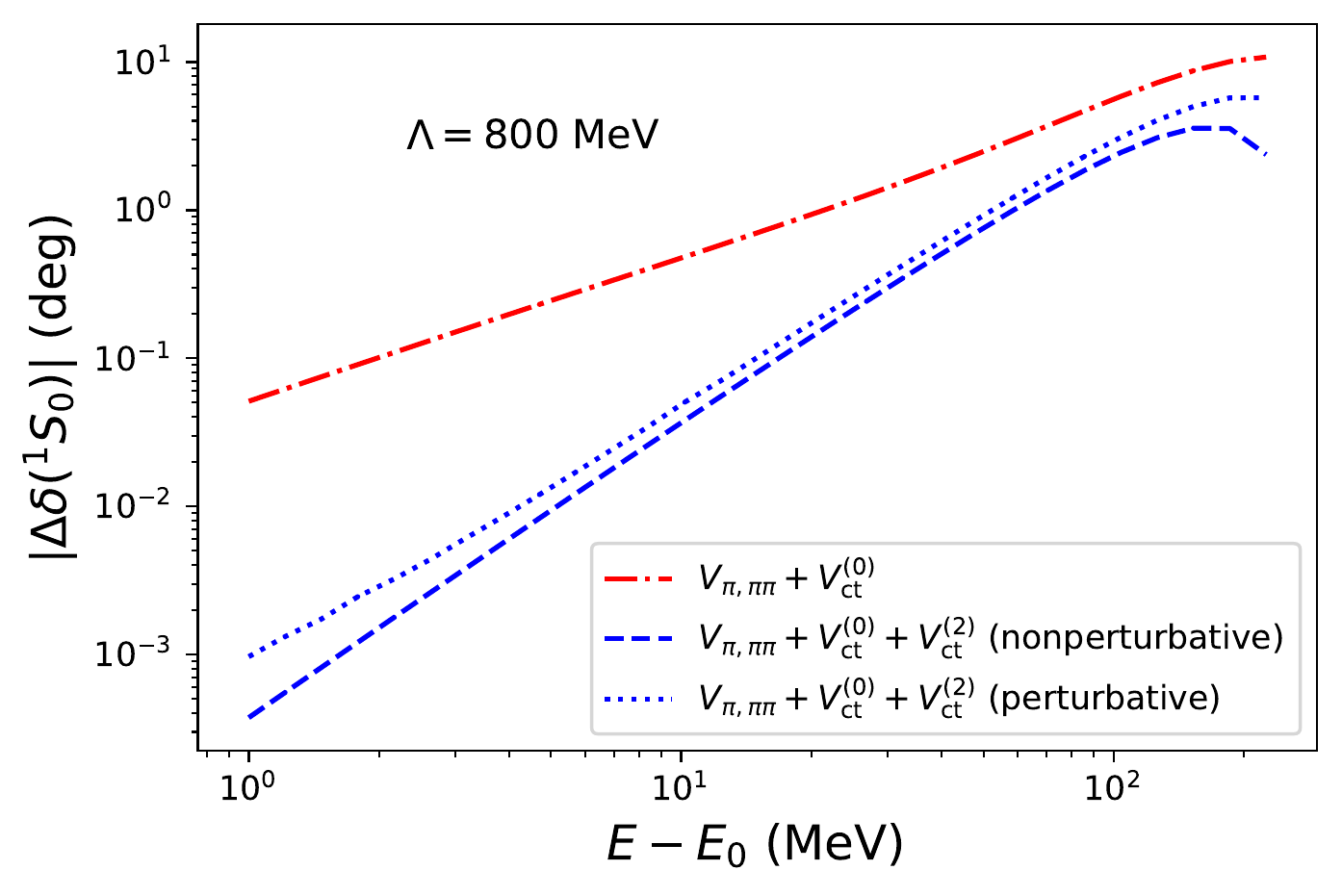}
  \caption{(Color online) Log-log plot of absolute differences to the reference  phase shifts versus the energy difference to the matching point. The cutoff is $800$~MeV. Results for the potentials $V_{\pi,\pi\pi}+V_{\rm ct}^{(0)}$ and $V_{\pi,\pi\pi}+V_{\rm ct}^{(0)}+V_{\rm ct}^{(2)}$ are shown as dash-dotted red and as blue lines, respectively. The dashed and the dotted blue line correspond to the  non-perturbative and the perturbative inclusion of the subleading contact $V_{\rm ct}^{(2)}$, respectively.}
  \label{fig:LOvNLO-Lepage}
\end{figure}

Let us discuss the non-perturbative approach. A simultaneous fit of the two LECs $(\tilde{C},C)$ to the phase shift and its slope at $E_0$ is somewhat challenging. Instead, we first calibrated $\tilde{C}$ and subsequently determined $C$ such that the reference phase shift is reproduced at $E_0$. Repeating this procedure for various values of $\tilde{C}$ yields a one-parameter curve $C(\tilde{C})$. Interestingly, we found that this is a quadratic function to very high accuracy. We then moved along this parabola and determined the point where the slope of the phase shift agrees with the reference as well. The resulting phase shifts are shown as  a dashed blue line in Fig.~\ref{fig:LOvNLO} and compared to our leading-order results (red dash-dotted line). The improvement is clearly visible.
Figure~\ref{fig:LOvNLO-Lepage} shows the absolute differences between our phase shifts and the reference 
on a log-log plot. The systematic power-law improvement from the quadratic contact is evident. We also note that $C\Lambda_{\rm b}^2/\tilde{C}\approx -4$, and this is consistent with expectations from naive dimensional analysis where this dimensionless number should be ${\cal O}(1)$ in size. 

In a second approach, we treated the contact~(\ref{contact2}) in perturbation theory. Our leading-order theory is adjusted to the reference phase shift at $E_0$ and the corresponding LECs are $(\tilde{C},C)=(\tilde{C}_0,0)$. In a perturbative approach the LECs become $(\tilde{C}_0+\delta\tilde{C},\delta C)$ in presence of the potential $V_{\rm ct}^{(2)}$. We expand the phase shift as 
\begin{align}
\label{phasePT}
  \delta(E) &\approx \delta(E){\rvert}_0
  + \frac{\partial \delta(E)}{\partial \tilde{C}}{\biggr\rvert}_0 \delta \tilde{C}
  + \frac{\partial \delta(E)}{\partial C}{\biggr\rvert}_0 \delta C
\end{align}
Here, the subscript $0$ implies that all functions are evaluated at $(\tilde{C}_0,0)$. We determine the derivatives numerically. Keeping the phase shift at $E_0$ unchanged implies 
\begin{align}
    \frac{\partial \delta(E_0)}{\partial \tilde{C}}{\biggr\rvert}_0 \delta\tilde{C} +
    \frac{\partial \delta(E_0)}{\partial C}{\biggr\rvert}_0 \delta C = 0 \ ,
\end{align}
and, thus, a linear relation between $\delta\tilde{C}$ and $\delta C$. A one-parameter search along this line yields the optimal point that also reproduces the slope of the reference phase shifts at $E_0$. We then compute the resulting phase shifts using Eq.~(\ref{phasePT}). The corresponding results are shown as blue dotted lines in Figs.~\ref{fig:LOvNLO} and \ref{fig:LOvNLO-Lepage}. We see that the perturbative and non-perturbative solutions phase shifts are close to each other, and that both yield a power-law improvement of our leading-order results.  We also note that $\Lambda_{\rm b}^2\delta C/(\tilde{C}_0+\delta C)\approx -0.03$, and this is smaller in magnitude than expected from naive dimensional analysis.

\begin{figure}[!htbp]
    \includegraphics[width=\linewidth]{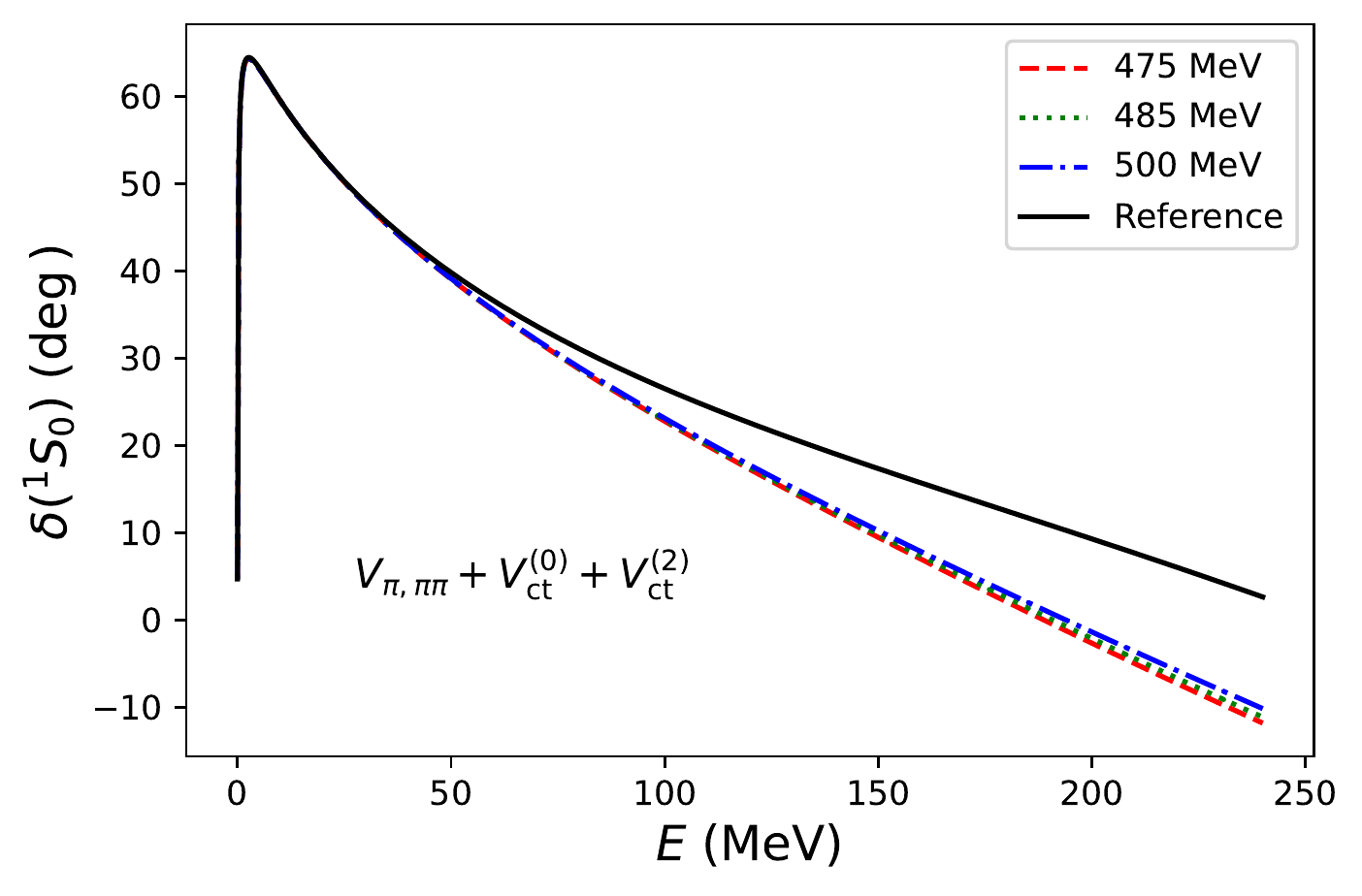}
    \caption{(Color online) The $^1S_0$ phase shifts as a function of the laboratory energy $E$ with the interaction $V_{\pi,\pi\pi}+V_{\rm ct}^{(0)} + V_{\rm ct}^{(2)}$ at low cutoffs as indicated and compared to the reference (solid black line). The additional contact $V_{\rm ct}^{(2)}$ restores the RG invariance that was lacking in Fig.~\ref{fig:small}.}
    \label{fig:smallNLO}
\end{figure}

We also revisited the low cutoffs at and below the scale $\Lambda_\text{TPE}$ in Eq.~(\ref{lam_tpe}) and employed the subleading contact $V_{\rm ct}^{(2)}$ in Eq.~(\ref{contact2}) non-perturbatively by matching the phase shift and its slope to the reference at a laboratory energy of 15~MeV. The  results are shown in Fig.~\ref{fig:smallNLO} and comparison with Fig.~\ref{fig:small} shows that RG invariance is restored also at low momentum cutoffs. 

\section{The role of $\Delta$ isobar degrees of freedom}
\label{sec:delta}

The strong contributions of the subleading TPE in chiral EFT, i.e., the relatively large values of the pion-nucleon LECs $c_i$,  are usually attributed to ``resonance saturation." Consequently, these couplings become more natural in size when the  $\Delta$ isobar degrees of freedom are included. Notably, such chiral EFTs have TPE terms that involve a $\Delta$ excitation already at NLO in the Weinberg power counting. The corresponding expressions for the potentials~(\ref{tpe}) were derived by~\textcite{kaiser1998} and we use those published by \textcite{krebs2007} in their Eqs.~(2.5) to (2.8). The chiral potential we employ thus consists of the OPE [given in Eq.~(\ref{ope})], leading TPE [given in Eq.~(\ref{nlo})], the contact~(\ref{contact0}), and the leading $\Delta$ contributions to TPE. In our numerical implementation this potential is regularized with $n=4$ in the regulator~(\ref{regulator}) and spectral-function regularization was used with a cutoff of $\tilde{\Lambda}=$700~MeV.

\begin{figure}[!htbp]
    \includegraphics[width=\linewidth]{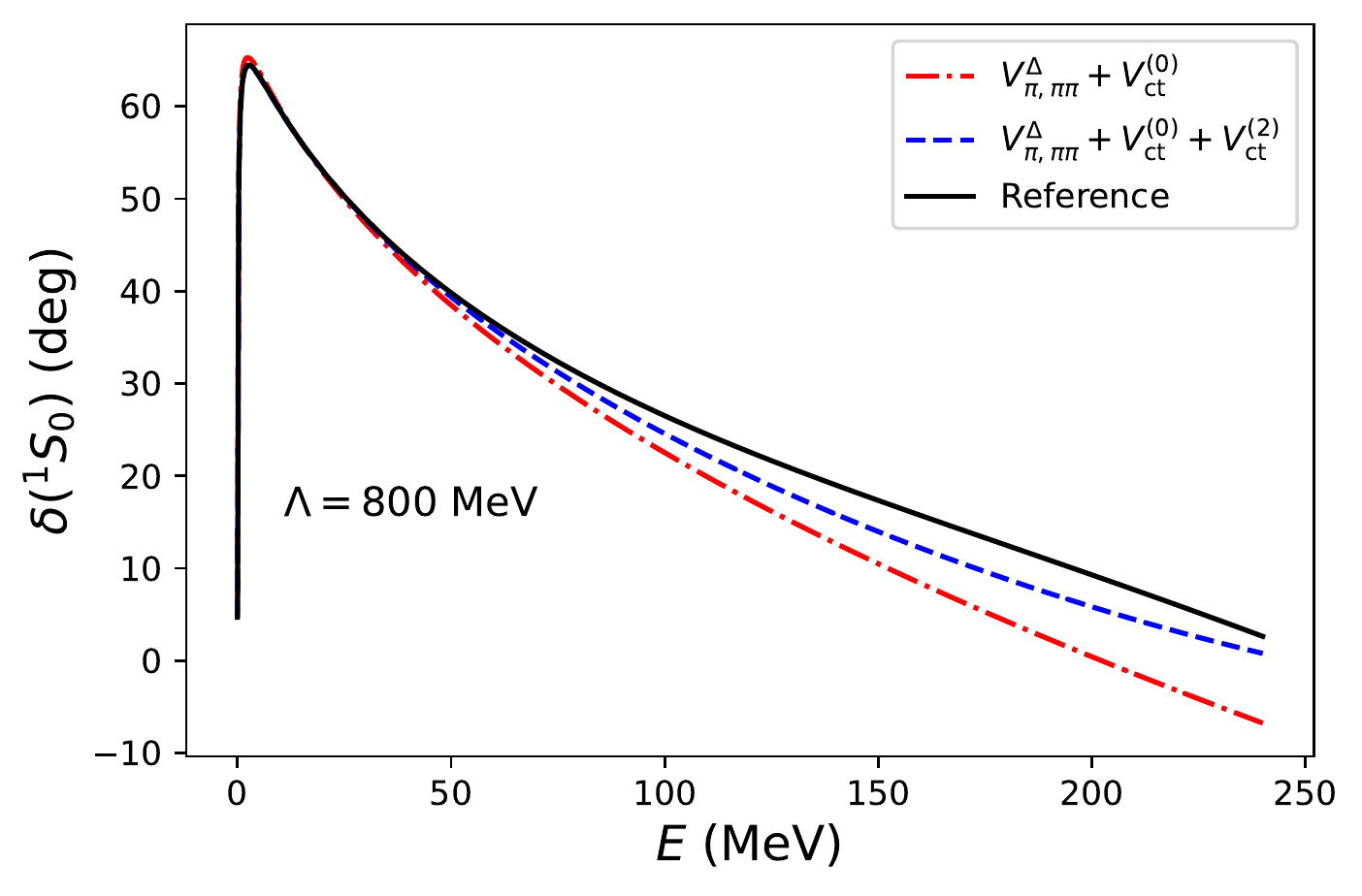}
    \caption{(Color online) The $^1S_0$ phase shifts as a function of the laboratory energy $E$ for the $V_{\pi,\pi\pi}^\Delta$ interaction, i.e., our leading-order potential with $\Delta$ isobars included in the leading TPE potential. The dash-dotted red line uses a single contact $V_{\rm ct}^{(0)}$ to reproduce the value of the reference phase shift at $E_0=15$~MeV. The dashed blue line shows the phase shift when leading and subleading contacts $V_{\rm ct}^{(0)} + V_{\rm ct}^{(2)}$ are fit to reproduce the value and the slope of the reference phase shift at $E_0$. The cutoff is $800$~MeV, and the reference phase shifts are shown as a solid black line. The results are close to those shown in Fig.~\ref{fig:LOvNLO}.}
    \label{fig:LOvNLODelta}
\end{figure}

We repeated the calculations presented above and found very similar results regarding the quality of the phase shifts in the $^1S_0$ partial wave, a breakdown momentum $\Lambda_\text{b}\approx 500$~MeV, and a systematic power-law improvement when the contact~(\ref{contact2}) quadratic in momenta is included. An example is shown in Fig.~\ref{fig:LOvNLODelta}, to be compared with Fig.~\ref{fig:LOvNLO}. 

\section{Summary and discussion}
\label{sec:summary}

We analyzed chiral EFT in the $^1S_0$ partial wave and propose to promote leading and subleading TPE to accompany OPE along with a single contact $V_{\rm ct}^{(0)}$ at leading order. Naturally, power counting in EFT applies to observables or amplitudes. Nevertheless, the promotion of TPE to leading order is inspired by the unexpectedly large matrix elements of the TPE potential that formally enter at NNLO in the Weinberg power counting. We find that our leading-order interaction cures two problems with the standard approach. First, the phase shifts in the $^1S_0$ partial wave are accurately reproduced and RG invariant for scattering energies of the order $m_\pi^2/m_N$ and for large cutoffs. Second, the estimated breakdown momentum $\Lambda_\text{b} \approx 500$ MeV in $^1S_0$ is consistent with assumptions from chiral EFT. We also showed that adding a contact quadratic in momenta as a subleading correction leads to a systematic power-law improvement of the phase shifts. We find that including the leading $\Delta$ contributions to TPE as a new leading-order contribution achieves the same. We expect that additional and smaller pion contributions and higher-order contacts to yield additional and systematic power-law improvements. We also pointed out that chiral EFTs with cutoffs below about $500$~MeV are really ``two-pion-less" EFTs as they cut off parts of the TPE; in such cases higher-order contacts are needed for maintaining  proper RG invariance.

\begin{acknowledgments}
We thank Daniel Phillips for insightful and useful discussions. We also thank the participants of the INT program ``Nuclear Forces for Precision Nuclear Physics (21-1b)'' for many useful exchanges and discussions. This material is based upon work supported by the U.S.\ Department of Energy, Office of Science, Office of Nuclear Physics under award numbers DE-FG02-96ER40963 and DE-SC0018223 (NUCLEI SciDAC-4 collaboration), and contract No. DE-AC05-00OR22725 with UT-Battelle, LLC (Oak Ridge National Laboratory), the Swedish Research Council grant number 2020-005127, the European Research Council (ERC) under the European Union's Horizon 2020 research and innovation programme (grant agreement number 758027), by the Deutsche Forschungsgemeinschaft (DFG, German Research Foundation) – Projektnummer 279384907 – CRC 1245. and by the National Science Foundation under Grant Nos. PHY-1555030 and PHY-2111426.  
\end{acknowledgments}


\begin{thebibliography}{59}%
\makeatletter
\providecommand \@ifxundefined [1]{%
 \@ifx{#1\undefined}
}%
\providecommand \@ifnum [1]{%
 \ifnum #1\expandafter \@firstoftwo
 \else \expandafter \@secondoftwo
 \fi
}%
\providecommand \@ifx [1]{%
 \ifx #1\expandafter \@firstoftwo
 \else \expandafter \@secondoftwo
 \fi
}%
\providecommand \natexlab [1]{#1}%
\providecommand \enquote  [1]{``#1''}%
\providecommand \bibnamefont  [1]{#1}%
\providecommand \bibfnamefont [1]{#1}%
\providecommand \citenamefont [1]{#1}%
\providecommand \href@noop [0]{\@secondoftwo}%
\providecommand \href [0]{\begingroup \@sanitize@url \@href}%
\providecommand \@href[1]{\@@startlink{#1}\@@href}%
\providecommand \@@href[1]{\endgroup#1\@@endlink}%
\providecommand \@sanitize@url [0]{\catcode `\\12\catcode `\$12\catcode
  `\&12\catcode `\#12\catcode `\^12\catcode `\_12\catcode `\%12\relax}%
\providecommand \@@startlink[1]{}%
\providecommand \@@endlink[0]{}%
\providecommand \url  [0]{\begingroup\@sanitize@url \@url }%
\providecommand \@url [1]{\endgroup\@href {#1}{\urlprefix }}%
\providecommand \urlprefix  [0]{URL }%
\providecommand \Eprint [0]{\href }%
\providecommand \doibase [0]{https://doi.org/}%
\providecommand \selectlanguage [0]{\@gobble}%
\providecommand \bibinfo  [0]{\@secondoftwo}%
\providecommand \bibfield  [0]{\@secondoftwo}%
\providecommand \translation [1]{[#1]}%
\providecommand \BibitemOpen [0]{}%
\providecommand \bibitemStop [0]{}%
\providecommand \bibitemNoStop [0]{.\EOS\space}%
\providecommand \EOS [0]{\spacefactor3000\relax}%
\providecommand \BibitemShut  [1]{\csname bibitem#1\endcsname}%
\let\auto@bib@innerbib\@empty
\bibitem [{\citenamefont {Yukawa}(1935)}]{yukawa1935}%
  \BibitemOpen
  \bibfield  {author} {\bibinfo {author} {\bibfnamefont {H.}~\bibnamefont
  {Yukawa}},\ }\bibfield  {title} {\bibinfo {title} {On the interaction of
  elementary particles. i},\ }\href
  {https://doi.org/10.11429/ppmsj1919.17.0_48} {\bibfield  {journal} {\bibinfo
  {journal} {Proceedings of the Physico-Mathematical Society of Japan. 3rd
  Series}\ }\textbf {\bibinfo {volume} {17}},\ \bibinfo {pages} {48} (\bibinfo
  {year} {1935})}\BibitemShut {NoStop}%
\bibitem [{\citenamefont {Weinberg}(1990)}]{weinberg1990}%
  \BibitemOpen
  \bibfield  {author} {\bibinfo {author} {\bibfnamefont {S.}~\bibnamefont
  {Weinberg}},\ }\bibfield  {title} {\bibinfo {title} {Nuclear forces from
  chiral lagrangians},\ }\href {https://doi.org/10.1016/0370-2693(90)90938-3}
  {\bibfield  {journal} {\bibinfo  {journal} {Phys. Lett. B}\ }\textbf
  {\bibinfo {volume} {251}},\ \bibinfo {pages} {288 } (\bibinfo {year}
  {1990})}\BibitemShut {NoStop}%
\bibitem [{\citenamefont {Weinberg}(1991)}]{weinberg1991}%
  \BibitemOpen
  \bibfield  {author} {\bibinfo {author} {\bibfnamefont {S.}~\bibnamefont
  {Weinberg}},\ }\bibfield  {title} {\bibinfo {title} {Effective chiral
  lagrangians for nucleon-pion interactions and nuclear forces},\ }\href
  {https://doi.org/10.1016/0550-3213(91)90231-L} {\bibfield  {journal}
  {\bibinfo  {journal} {Nuclear Physics B}\ }\textbf {\bibinfo {volume}
  {363}},\ \bibinfo {pages} {3 } (\bibinfo {year} {1991})}\BibitemShut
  {NoStop}%
\bibitem [{\citenamefont {Ord{\'o}{\~n}ez}\ and\ \citenamefont {van
  Kolck}(1992)}]{ordonez1992}%
  \BibitemOpen
  \bibfield  {author} {\bibinfo {author} {\bibfnamefont {C.}~\bibnamefont
  {Ord{\'o}{\~n}ez}}\ and\ \bibinfo {author} {\bibfnamefont {U.}~\bibnamefont
  {van Kolck}},\ }\bibfield  {title} {\bibinfo {title} {Chiral lagrangians and
  nuclear forces},\ }\href {https://doi.org/10.1016/0370-2693(92)91404-W}
  {\bibfield  {journal} {\bibinfo  {journal} {Phys. Lett. B}\ }\textbf
  {\bibinfo {volume} {291}},\ \bibinfo {pages} {459 } (\bibinfo {year}
  {1992})}\BibitemShut {NoStop}%
\bibitem [{\citenamefont {van Kolck}(1994)}]{vankolck1994}%
  \BibitemOpen
  \bibfield  {author} {\bibinfo {author} {\bibfnamefont {U.}~\bibnamefont {van
  Kolck}},\ }\bibfield  {title} {\bibinfo {title} {{Few-nucleon forces from
  chiral Lagrangians}},\ }\href {https://doi.org/10.1103/PhysRevC.49.2932}
  {\bibfield  {journal} {\bibinfo  {journal} {Phys. Rev. C}\ }\textbf {\bibinfo
  {volume} {49}},\ \bibinfo {pages} {2932} (\bibinfo {year}
  {1994})}\BibitemShut {NoStop}%
\bibitem [{\citenamefont {Kaiser}\ \emph {et~al.}(1997)\citenamefont {Kaiser},
  \citenamefont {Brockmann},\ and\ \citenamefont {Weise}}]{kaiser1997}%
  \BibitemOpen
  \bibfield  {author} {\bibinfo {author} {\bibfnamefont {N.}~\bibnamefont
  {Kaiser}}, \bibinfo {author} {\bibfnamefont {R.}~\bibnamefont {Brockmann}},\
  and\ \bibinfo {author} {\bibfnamefont {W.}~\bibnamefont {Weise}},\ }\bibfield
   {title} {\bibinfo {title} {Peripheral nucleon-nucleon phase shifts and
  chiral symmetry},\ }\href {https://doi.org/10.1016/S0375-9474(97)00586-1}
  {\bibfield  {journal} {\bibinfo  {journal} {Nucl. Phys. A}\ }\textbf
  {\bibinfo {volume} {625}},\ \bibinfo {pages} {758 } (\bibinfo {year}
  {1997})}\BibitemShut {NoStop}%
\bibitem [{\citenamefont {Epelbaum}\ \emph {et~al.}(2000)\citenamefont
  {Epelbaum}, \citenamefont {Gl{\"o}ckle},\ and\ \citenamefont
  {Mei{\ss}ner}}]{epelbaum2000}%
  \BibitemOpen
  \bibfield  {author} {\bibinfo {author} {\bibfnamefont {E.}~\bibnamefont
  {Epelbaum}}, \bibinfo {author} {\bibfnamefont {W.}~\bibnamefont
  {Gl{\"o}ckle}},\ and\ \bibinfo {author} {\bibfnamefont {U.-G.}\ \bibnamefont
  {Mei{\ss}ner}},\ }\bibfield  {title} {\bibinfo {title} {Nuclear forces from
  chiral lagrangians using the method of unitary transformation ii: The
  two-nucleon system},\ }\href {https://doi.org/10.1016/S0375-9474(99)00821-0}
  {\bibfield  {journal} {\bibinfo  {journal} {Nucl. Phys. A}\ }\textbf
  {\bibinfo {volume} {671}},\ \bibinfo {pages} {295 } (\bibinfo {year}
  {2000})}\BibitemShut {NoStop}%
\bibitem [{\citenamefont {Epelbaum}\ \emph {et~al.}(2009)\citenamefont
  {Epelbaum}, \citenamefont {Hammer},\ and\ \citenamefont
  {Mei\ss{}ner}}]{epelbaum2009}%
  \BibitemOpen
  \bibfield  {author} {\bibinfo {author} {\bibfnamefont {E.}~\bibnamefont
  {Epelbaum}}, \bibinfo {author} {\bibfnamefont {H.-W.}\ \bibnamefont
  {Hammer}},\ and\ \bibinfo {author} {\bibfnamefont {U.-G.}\ \bibnamefont
  {Mei\ss{}ner}},\ }\bibfield  {title} {\bibinfo {title} {Modern theory of
  nuclear forces},\ }\href {https://doi.org/10.1103/RevModPhys.81.1773}
  {\bibfield  {journal} {\bibinfo  {journal} {Rev. Mod. Phys.}\ }\textbf
  {\bibinfo {volume} {81}},\ \bibinfo {pages} {1773} (\bibinfo {year}
  {2009})}\BibitemShut {NoStop}%
\bibitem [{\citenamefont {Hammer}\ \emph {et~al.}(2020)\citenamefont {Hammer},
  \citenamefont {K\"onig},\ and\ \citenamefont {van Kolck}}]{Hammer:2019poc}%
  \BibitemOpen
  \bibfield  {author} {\bibinfo {author} {\bibfnamefont {H.~W.}\ \bibnamefont
  {Hammer}}, \bibinfo {author} {\bibfnamefont {S.}~\bibnamefont {K\"onig}},\
  and\ \bibinfo {author} {\bibfnamefont {U.}~\bibnamefont {van Kolck}},\
  }\bibfield  {title} {\bibinfo {title} {{Nuclear effective field theory:
  status and perspectives}},\ }\href
  {https://doi.org/10.1103/RevModPhys.92.025004} {\bibfield  {journal}
  {\bibinfo  {journal} {Rev. Mod. Phys.}\ }\textbf {\bibinfo {volume} {92}},\
  \bibinfo {pages} {025004} (\bibinfo {year} {2020})},\ \Eprint
  {https://arxiv.org/abs/1906.12122} {arXiv:1906.12122 [nucl-th]} \BibitemShut
  {NoStop}%
\bibitem [{\citenamefont {Reinert}\ \emph {et~al.}(2018)\citenamefont
  {Reinert}, \citenamefont {Krebs},\ and\ \citenamefont
  {Epelbaum}}]{reinert2018}%
  \BibitemOpen
  \bibfield  {author} {\bibinfo {author} {\bibfnamefont {P.}~\bibnamefont
  {Reinert}}, \bibinfo {author} {\bibfnamefont {H.}~\bibnamefont {Krebs}},\
  and\ \bibinfo {author} {\bibfnamefont {E.}~\bibnamefont {Epelbaum}},\
  }\bibfield  {title} {\bibinfo {title} {Semilocal momentum-space regularized
  chiral two-nucleon potentials up to fifth order},\ }\href
  {https://doi.org/10.1140/epja/i2018-12516-4} {\bibfield  {journal} {\bibinfo
  {journal} {The European Physical Journal A}\ }\textbf {\bibinfo {volume}
  {54}},\ \bibinfo {pages} {86} (\bibinfo {year} {2018})}\BibitemShut {NoStop}%
\bibitem [{\citenamefont {Wesolowski}\ \emph {et~al.}(2019)\citenamefont
  {Wesolowski}, \citenamefont {Furnstahl}, \citenamefont {Melendez},\ and\
  \citenamefont {Phillips}}]{wesolowski2019}%
  \BibitemOpen
  \bibfield  {author} {\bibinfo {author} {\bibfnamefont {S.}~\bibnamefont
  {Wesolowski}}, \bibinfo {author} {\bibfnamefont {R.~J.}\ \bibnamefont
  {Furnstahl}}, \bibinfo {author} {\bibfnamefont {J.~A.}\ \bibnamefont
  {Melendez}},\ and\ \bibinfo {author} {\bibfnamefont {D.~R.}\ \bibnamefont
  {Phillips}},\ }\bibfield  {title} {\bibinfo {title} {Exploring bayesian
  parameter estimation for chiral effective field theory using
  nucleon{\textendash}nucleon phase shifts},\ }\href
  {https://doi.org/10.1088/1361-6471/aaf5fc} {\bibfield  {journal} {\bibinfo
  {journal} {Journal of Physics G: Nuclear and Particle Physics}\ }\textbf
  {\bibinfo {volume} {46}},\ \bibinfo {pages} {045102} (\bibinfo {year}
  {2019})}\BibitemShut {NoStop}%
\bibitem [{\citenamefont {{Furnstahl}}\ \emph {et~al.}(2021)\citenamefont
  {{Furnstahl}}, \citenamefont {{Hammer}},\ and\ \citenamefont
  {{Schwenk}}}]{furnstahl2021}%
  \BibitemOpen
  \bibfield  {author} {\bibinfo {author} {\bibfnamefont {R.~J.}\ \bibnamefont
  {{Furnstahl}}}, \bibinfo {author} {\bibfnamefont {H.~W.}\ \bibnamefont
  {{Hammer}}},\ and\ \bibinfo {author} {\bibfnamefont {A.}~\bibnamefont
  {{Schwenk}}},\ }\bibfield  {title} {\bibinfo {title} {{Nuclear Structure at
  the Crossroads}},\ }\href {https://doi.org/10.1007/s00601-021-01658-5}
  {\bibfield  {journal} {\bibinfo  {journal} {Few-Body Systems}\ }\textbf
  {\bibinfo {volume} {62}},\ \bibinfo {pages} {72} (\bibinfo {year}
  {2021})}\BibitemShut {NoStop}%
\bibitem [{\citenamefont {Carlsson}\ \emph {et~al.}(2016)\citenamefont
  {Carlsson}, \citenamefont {Ekstr\"om}, \citenamefont {Forss\'en},
  \citenamefont {Str\"omberg}, \citenamefont {Jansen}, \citenamefont {Lilja},
  \citenamefont {Lindby}, \citenamefont {Mattsson},\ and\ \citenamefont
  {Wendt}}]{carlsson2016}%
  \BibitemOpen
  \bibfield  {author} {\bibinfo {author} {\bibfnamefont {B.~D.}\ \bibnamefont
  {Carlsson}}, \bibinfo {author} {\bibfnamefont {A.}~\bibnamefont {Ekstr\"om}},
  \bibinfo {author} {\bibfnamefont {C.}~\bibnamefont {Forss\'en}}, \bibinfo
  {author} {\bibfnamefont {D.~F.}\ \bibnamefont {Str\"omberg}}, \bibinfo
  {author} {\bibfnamefont {G.~R.}\ \bibnamefont {Jansen}}, \bibinfo {author}
  {\bibfnamefont {O.}~\bibnamefont {Lilja}}, \bibinfo {author} {\bibfnamefont
  {M.}~\bibnamefont {Lindby}}, \bibinfo {author} {\bibfnamefont {B.~A.}\
  \bibnamefont {Mattsson}},\ and\ \bibinfo {author} {\bibfnamefont {K.~A.}\
  \bibnamefont {Wendt}},\ }\bibfield  {title} {\bibinfo {title} {Uncertainty
  analysis and order-by-order optimization of chiral nuclear interactions},\
  }\href {https://doi.org/10.1103/PhysRevX.6.011019} {\bibfield  {journal}
  {\bibinfo  {journal} {Phys. Rev. X}\ }\textbf {\bibinfo {volume} {6}},\
  \bibinfo {pages} {011019} (\bibinfo {year} {2016})}\BibitemShut {NoStop}%
\bibitem [{\citenamefont {Yang}\ \emph {et~al.}(2021)\citenamefont {Yang},
  \citenamefont {Ekstr\"om}, \citenamefont {Forss\'en},\ and\ \citenamefont
  {Hagen}}]{yang2021}%
  \BibitemOpen
  \bibfield  {author} {\bibinfo {author} {\bibfnamefont {C.-J.}\ \bibnamefont
  {Yang}}, \bibinfo {author} {\bibfnamefont {A.}~\bibnamefont {Ekstr\"om}},
  \bibinfo {author} {\bibfnamefont {C.}~\bibnamefont {Forss\'en}},\ and\
  \bibinfo {author} {\bibfnamefont {G.}~\bibnamefont {Hagen}},\ }\bibfield
  {title} {\bibinfo {title} {Power counting in chiral effective field theory
  and nuclear binding},\ }\href {https://doi.org/10.1103/PhysRevC.103.054304}
  {\bibfield  {journal} {\bibinfo  {journal} {Phys. Rev. C}\ }\textbf {\bibinfo
  {volume} {103}},\ \bibinfo {pages} {054304} (\bibinfo {year}
  {2021})}\BibitemShut {NoStop}%
\bibitem [{\citenamefont {Maris}\ \emph {et~al.}(2021)\citenamefont {Maris},
  \citenamefont {Epelbaum}, \citenamefont {Furnstahl}, \citenamefont {Golak},
  \citenamefont {Hebeler}, \citenamefont {H\"uther}, \citenamefont {Kamada},
  \citenamefont {Krebs}, \citenamefont {Mei\ss{}ner}, \citenamefont {Melendez},
  \citenamefont {Nogga}, \citenamefont {Reinert}, \citenamefont {Roth},
  \citenamefont {Skibi\ifmmode~\acute{n}\else \'{n}\fi{}ski}, \citenamefont
  {Soloviov}, \citenamefont {Topolnicki}, \citenamefont {Vary}, \citenamefont
  {Volkotrub}, \citenamefont {Wita\l{}a},\ and\ \citenamefont
  {Wolfgruber}}]{maris2021}%
  \BibitemOpen
  \bibfield  {author} {\bibinfo {author} {\bibfnamefont {P.}~\bibnamefont
  {Maris}}, \bibinfo {author} {\bibfnamefont {E.}~\bibnamefont {Epelbaum}},
  \bibinfo {author} {\bibfnamefont {R.~J.}\ \bibnamefont {Furnstahl}}, \bibinfo
  {author} {\bibfnamefont {J.}~\bibnamefont {Golak}}, \bibinfo {author}
  {\bibfnamefont {K.}~\bibnamefont {Hebeler}}, \bibinfo {author} {\bibfnamefont
  {T.}~\bibnamefont {H\"uther}}, \bibinfo {author} {\bibfnamefont
  {H.}~\bibnamefont {Kamada}}, \bibinfo {author} {\bibfnamefont
  {H.}~\bibnamefont {Krebs}}, \bibinfo {author} {\bibfnamefont {U.-G.}\
  \bibnamefont {Mei\ss{}ner}}, \bibinfo {author} {\bibfnamefont {J.~A.}\
  \bibnamefont {Melendez}}, \bibinfo {author} {\bibfnamefont {A.}~\bibnamefont
  {Nogga}}, \bibinfo {author} {\bibfnamefont {P.}~\bibnamefont {Reinert}},
  \bibinfo {author} {\bibfnamefont {R.}~\bibnamefont {Roth}}, \bibinfo {author}
  {\bibfnamefont {R.}~\bibnamefont {Skibi\ifmmode~\acute{n}\else
  \'{n}\fi{}ski}}, \bibinfo {author} {\bibfnamefont {V.}~\bibnamefont
  {Soloviov}}, \bibinfo {author} {\bibfnamefont {K.}~\bibnamefont
  {Topolnicki}}, \bibinfo {author} {\bibfnamefont {J.~P.}\ \bibnamefont
  {Vary}}, \bibinfo {author} {\bibfnamefont {Y.}~\bibnamefont {Volkotrub}},
  \bibinfo {author} {\bibfnamefont {H.}~\bibnamefont {Wita\l{}a}},\ and\
  \bibinfo {author} {\bibfnamefont {T.}~\bibnamefont {Wolfgruber}} (\bibinfo
  {collaboration} {LENPIC Collaboration}),\ }\bibfield  {title} {\bibinfo
  {title} {Light nuclei with semilocal momentum-space regularized chiral
  interactions up to third order},\ }\href
  {https://doi.org/10.1103/PhysRevC.103.054001} {\bibfield  {journal} {\bibinfo
   {journal} {Phys. Rev. C}\ }\textbf {\bibinfo {volume} {103}},\ \bibinfo
  {pages} {054001} (\bibinfo {year} {2021})}\BibitemShut {NoStop}%
\bibitem [{\citenamefont {Entem}\ and\ \citenamefont
  {Machleidt}(2003)}]{entem2003}%
  \BibitemOpen
  \bibfield  {author} {\bibinfo {author} {\bibfnamefont {D.~R.}\ \bibnamefont
  {Entem}}\ and\ \bibinfo {author} {\bibfnamefont {R.}~\bibnamefont
  {Machleidt}},\ }\bibfield  {title} {\bibinfo {title} {Accurate
  charge-dependent nucleon-nucleon potential at fourth order of chiral
  perturbation theory},\ }\href {https://doi.org/10.1103/PhysRevC.68.041001}
  {\bibfield  {journal} {\bibinfo  {journal} {Phys. Rev. C}\ }\textbf {\bibinfo
  {volume} {68}},\ \bibinfo {pages} {041001} (\bibinfo {year}
  {2003})}\BibitemShut {NoStop}%
\bibitem [{\citenamefont {Kaplan}\ \emph {et~al.}(1998)\citenamefont {Kaplan},
  \citenamefont {Savage},\ and\ \citenamefont {Wise}}]{kaplan1998}%
  \BibitemOpen
  \bibfield  {author} {\bibinfo {author} {\bibfnamefont {D.~B.}\ \bibnamefont
  {Kaplan}}, \bibinfo {author} {\bibfnamefont {M.~J.}\ \bibnamefont {Savage}},\
  and\ \bibinfo {author} {\bibfnamefont {M.~B.}\ \bibnamefont {Wise}},\
  }\bibfield  {title} {\bibinfo {title} {A new expansion for nucleon-nucleon
  interactions},\ }\href {https://doi.org/10.1016/S0370-2693(98)00210-X}
  {\bibfield  {journal} {\bibinfo  {journal} {Physics Letters B}\ }\textbf
  {\bibinfo {volume} {424}},\ \bibinfo {pages} {390 } (\bibinfo {year}
  {1998})}\BibitemShut {NoStop}%
\bibitem [{\citenamefont {Frederico}\ \emph {et~al.}(1999)\citenamefont
  {Frederico}, \citenamefont {Timoteo},\ and\ \citenamefont
  {Tomio}}]{frederico1999}%
  \BibitemOpen
  \bibfield  {author} {\bibinfo {author} {\bibfnamefont {T.}~\bibnamefont
  {Frederico}}, \bibinfo {author} {\bibfnamefont {V.~S.}\ \bibnamefont
  {Timoteo}},\ and\ \bibinfo {author} {\bibfnamefont {L.}~\bibnamefont
  {Tomio}},\ }\bibfield  {title} {\bibinfo {title} {{Renormalization of the one
  pion exchange interaction}},\ }\href
  {https://doi.org/10.1016/S0375-9474(99)00234-1} {\bibfield  {journal}
  {\bibinfo  {journal} {Nucl. Phys. A}\ }\textbf {\bibinfo {volume} {653}},\
  \bibinfo {pages} {209} (\bibinfo {year} {1999})},\ \Eprint
  {https://arxiv.org/abs/nucl-th/9902052} {arXiv:nucl-th/9902052} \BibitemShut
  {NoStop}%
\bibitem [{\citenamefont {Nogga}\ \emph {et~al.}(2005)\citenamefont {Nogga},
  \citenamefont {Timmermans},\ and\ \citenamefont {Kolck}}]{nogga2005}%
  \BibitemOpen
  \bibfield  {author} {\bibinfo {author} {\bibfnamefont {A.}~\bibnamefont
  {Nogga}}, \bibinfo {author} {\bibfnamefont {R.~G.~E.}\ \bibnamefont
  {Timmermans}},\ and\ \bibinfo {author} {\bibfnamefont {U.~v.}\ \bibnamefont
  {Kolck}},\ }\bibfield  {title} {\bibinfo {title} {Renormalization of one-pion
  exchange and power counting},\ }\href
  {https://doi.org/10.1103/PhysRevC.72.054006} {\bibfield  {journal} {\bibinfo
  {journal} {Phys. Rev. C}\ }\textbf {\bibinfo {volume} {72}},\ \bibinfo
  {pages} {054006} (\bibinfo {year} {2005})}\BibitemShut {NoStop}%
\bibitem [{\citenamefont {Valderrama}\ and\ \citenamefont
  {Arriola}(2005)}]{valderrama2005}%
  \BibitemOpen
  \bibfield  {author} {\bibinfo {author} {\bibfnamefont {M.~P.}\ \bibnamefont
  {Valderrama}}\ and\ \bibinfo {author} {\bibfnamefont {E.~R.}\ \bibnamefont
  {Arriola}},\ }\bibfield  {title} {\bibinfo {title} {Renormalization of the
  deuteron with one pion exchange},\ }\href
  {https://doi.org/10.1103/PhysRevC.72.054002} {\bibfield  {journal} {\bibinfo
  {journal} {Phys. Rev. C}\ }\textbf {\bibinfo {volume} {72}},\ \bibinfo
  {pages} {054002} (\bibinfo {year} {2005})}\BibitemShut {NoStop}%
\bibitem [{\citenamefont {Pav\'on~Valderrama}\ and\ \citenamefont
  {Arriola}(2006)}]{valderrama2006}%
  \BibitemOpen
  \bibfield  {author} {\bibinfo {author} {\bibfnamefont {M.}~\bibnamefont
  {Pav\'on~Valderrama}}\ and\ \bibinfo {author} {\bibfnamefont {E.~R.}\
  \bibnamefont {Arriola}},\ }\bibfield  {title} {\bibinfo {title}
  {Renormalization of the $\mathit{NN}$ interaction with a chiral
  two-pion-exchange potential: Central phases and the deuteron},\ }\href
  {https://doi.org/10.1103/PhysRevC.74.054001} {\bibfield  {journal} {\bibinfo
  {journal} {Phys. Rev. C}\ }\textbf {\bibinfo {volume} {74}},\ \bibinfo
  {pages} {054001} (\bibinfo {year} {2006})}\BibitemShut {NoStop}%
\bibitem [{\citenamefont {Birse}(2006)}]{birse2006}%
  \BibitemOpen
  \bibfield  {author} {\bibinfo {author} {\bibfnamefont {M.~C.}\ \bibnamefont
  {Birse}},\ }\bibfield  {title} {\bibinfo {title} {Power counting with
  one-pion exchange},\ }\href {https://doi.org/10.1103/PhysRevC.74.014003}
  {\bibfield  {journal} {\bibinfo  {journal} {Phys. Rev. C}\ }\textbf {\bibinfo
  {volume} {74}},\ \bibinfo {pages} {014003} (\bibinfo {year}
  {2006})}\BibitemShut {NoStop}%
\bibitem [{\citenamefont {Shukla}\ \emph {et~al.}(2008)\citenamefont {Shukla},
  \citenamefont {Phillips},\ and\ \citenamefont {Mortenson}}]{shukla2008}%
  \BibitemOpen
  \bibfield  {author} {\bibinfo {author} {\bibfnamefont {D.}~\bibnamefont
  {Shukla}}, \bibinfo {author} {\bibfnamefont {D.~R.}\ \bibnamefont
  {Phillips}},\ and\ \bibinfo {author} {\bibfnamefont {E.}~\bibnamefont
  {Mortenson}},\ }\bibfield  {title} {\bibinfo {title} {Chiral potentials,
  perturbation theory and the $^1s_0$ channel of {NN} scattering},\ }\href
  {https://doi.org/10.1088/0954-3899/35/11/115009} {\bibfield  {journal}
  {\bibinfo  {journal} {Journal of Physics G: Nuclear and Particle Physics}\
  }\textbf {\bibinfo {volume} {35}},\ \bibinfo {pages} {115009} (\bibinfo
  {year} {2008})}\BibitemShut {NoStop}%
\bibitem [{\citenamefont {Yang}\ \emph {et~al.}(2008)\citenamefont {Yang},
  \citenamefont {Elster},\ and\ \citenamefont {Phillips}}]{yang2008}%
  \BibitemOpen
  \bibfield  {author} {\bibinfo {author} {\bibfnamefont {C.-J.}\ \bibnamefont
  {Yang}}, \bibinfo {author} {\bibfnamefont {C.}~\bibnamefont {Elster}},\ and\
  \bibinfo {author} {\bibfnamefont {D.~R.}\ \bibnamefont {Phillips}},\
  }\bibfield  {title} {\bibinfo {title} {Subtractive renormalization of the
  $\mathit{NN}$ scattering amplitude at leading order in chiral effective
  theory},\ }\href {https://doi.org/10.1103/PhysRevC.77.014002} {\bibfield
  {journal} {\bibinfo  {journal} {Phys. Rev. C}\ }\textbf {\bibinfo {volume}
  {77}},\ \bibinfo {pages} {014002} (\bibinfo {year} {2008})}\BibitemShut
  {NoStop}%
\bibitem [{\citenamefont {{Birse}}(2010)}]{birse2010}%
  \BibitemOpen
  \bibfield  {author} {\bibinfo {author} {\bibfnamefont {M.~C.}\ \bibnamefont
  {{Birse}}},\ }\bibfield  {title} {\bibinfo {title} {{Deconstructing
  $^{1}$S$_{0}$ nucleon-nucleon scattering}},\ }\href
  {https://doi.org/10.1140/epja/i2010-11034-9} {\bibfield  {journal} {\bibinfo
  {journal} {European Physical Journal A}\ }\textbf {\bibinfo {volume} {46}},\
  \bibinfo {pages} {231} (\bibinfo {year} {2010})}\BibitemShut {NoStop}%
\bibitem [{\citenamefont {Long}\ and\ \citenamefont
  {Yang}(2012{\natexlab{a}})}]{long2012}%
  \BibitemOpen
  \bibfield  {author} {\bibinfo {author} {\bibfnamefont {B.}~\bibnamefont
  {Long}}\ and\ \bibinfo {author} {\bibfnamefont {C.-J.}\ \bibnamefont
  {Yang}},\ }\bibfield  {title} {\bibinfo {title} {Short-range nuclear forces
  in singlet channels},\ }\href {https://doi.org/10.1103/PhysRevC.86.024001}
  {\bibfield  {journal} {\bibinfo  {journal} {Phys. Rev. C}\ }\textbf {\bibinfo
  {volume} {86}},\ \bibinfo {pages} {024001} (\bibinfo {year}
  {2012}{\natexlab{a}})}\BibitemShut {NoStop}%
\bibitem [{\citenamefont {Long}\ and\ \citenamefont
  {Yang}(2012{\natexlab{b}})}]{long2012b}%
  \BibitemOpen
  \bibfield  {author} {\bibinfo {author} {\bibfnamefont {B.}~\bibnamefont
  {Long}}\ and\ \bibinfo {author} {\bibfnamefont {C.-J.}\ \bibnamefont
  {Yang}},\ }\bibfield  {title} {\bibinfo {title} {Renormalizing chiral nuclear
  forces: Triplet channels},\ }\href
  {https://doi.org/10.1103/PhysRevC.85.034002} {\bibfield  {journal} {\bibinfo
  {journal} {Phys. Rev. C}\ }\textbf {\bibinfo {volume} {85}},\ \bibinfo
  {pages} {034002} (\bibinfo {year} {2012}{\natexlab{b}})}\BibitemShut
  {NoStop}%
\bibitem [{\citenamefont {Valderrama}\ and\ \citenamefont
  {Phillips}(2015)}]{valderrama2015}%
  \BibitemOpen
  \bibfield  {author} {\bibinfo {author} {\bibfnamefont {M.~P.}\ \bibnamefont
  {Valderrama}}\ and\ \bibinfo {author} {\bibfnamefont {D.~R.}\ \bibnamefont
  {Phillips}},\ }\bibfield  {title} {\bibinfo {title} {Power counting of
  contact-range currents in effective field theory},\ }\href
  {https://doi.org/10.1103/PhysRevLett.114.082502} {\bibfield  {journal}
  {\bibinfo  {journal} {Phys. Rev. Lett.}\ }\textbf {\bibinfo {volume} {114}},\
  \bibinfo {pages} {082502} (\bibinfo {year} {2015})}\BibitemShut {NoStop}%
\bibitem [{\citenamefont {S\'anchez~S\'anchez}\ \emph
  {et~al.}(2018)\citenamefont {S\'anchez~S\'anchez}, \citenamefont {Yang},
  \citenamefont {Long},\ and\ \citenamefont {van Kolck}}]{sanchez2018}%
  \BibitemOpen
  \bibfield  {author} {\bibinfo {author} {\bibfnamefont {M.}~\bibnamefont
  {S\'anchez~S\'anchez}}, \bibinfo {author} {\bibfnamefont {C.-J.}\
  \bibnamefont {Yang}}, \bibinfo {author} {\bibfnamefont {B.}~\bibnamefont
  {Long}},\ and\ \bibinfo {author} {\bibfnamefont {U.}~\bibnamefont {van
  Kolck}},\ }\bibfield  {title} {\bibinfo {title} {Two-nucleon $^{1}s_{0}$
  amplitude zero in chiral effective field theory},\ }\href
  {https://doi.org/10.1103/PhysRevC.97.024001} {\bibfield  {journal} {\bibinfo
  {journal} {Phys. Rev. C}\ }\textbf {\bibinfo {volume} {97}},\ \bibinfo
  {pages} {024001} (\bibinfo {year} {2018})}\BibitemShut {NoStop}%
\bibitem [{\citenamefont {Odell}\ \emph {et~al.}(2019)\citenamefont {Odell},
  \citenamefont {Deltuva}, \citenamefont {Bonilla},\ and\ \citenamefont
  {Platter}}]{odell2019}%
  \BibitemOpen
  \bibfield  {author} {\bibinfo {author} {\bibfnamefont {D.}~\bibnamefont
  {Odell}}, \bibinfo {author} {\bibfnamefont {A.}~\bibnamefont {Deltuva}},
  \bibinfo {author} {\bibfnamefont {J.}~\bibnamefont {Bonilla}},\ and\ \bibinfo
  {author} {\bibfnamefont {L.}~\bibnamefont {Platter}},\ }\bibfield  {title}
  {\bibinfo {title} {Renormalization of a finite-range inverse-cube
  potential},\ }\href {https://doi.org/10.1103/PhysRevC.100.054001} {\bibfield
  {journal} {\bibinfo  {journal} {Phys. Rev. C}\ }\textbf {\bibinfo {volume}
  {100}},\ \bibinfo {pages} {054001} (\bibinfo {year} {2019})}\BibitemShut
  {NoStop}%
\bibitem [{\citenamefont {S\'anchez~S\'anchez}\ \emph
  {et~al.}(2020)\citenamefont {S\'anchez~S\'anchez}, \citenamefont {Smirnova},
  \citenamefont {Shirokov}, \citenamefont {Maris},\ and\ \citenamefont
  {Vary}}]{sanchez2020}%
  \BibitemOpen
  \bibfield  {author} {\bibinfo {author} {\bibfnamefont {M.}~\bibnamefont
  {S\'anchez~S\'anchez}}, \bibinfo {author} {\bibfnamefont {N.~A.}\
  \bibnamefont {Smirnova}}, \bibinfo {author} {\bibfnamefont {A.~M.}\
  \bibnamefont {Shirokov}}, \bibinfo {author} {\bibfnamefont {P.}~\bibnamefont
  {Maris}},\ and\ \bibinfo {author} {\bibfnamefont {J.~P.}\ \bibnamefont
  {Vary}},\ }\bibfield  {title} {\bibinfo {title} {Improved description of
  light nuclei through chiral effective field theory at leading order},\ }\href
  {https://doi.org/10.1103/PhysRevC.102.024324} {\bibfield  {journal} {\bibinfo
   {journal} {Phys. Rev. C}\ }\textbf {\bibinfo {volume} {102}},\ \bibinfo
  {pages} {024324} (\bibinfo {year} {2020})}\BibitemShut {NoStop}%
\bibitem [{\citenamefont {van Kolck}(2020)}]{vankolck2020}%
  \BibitemOpen
  \bibfield  {author} {\bibinfo {author} {\bibfnamefont {U.}~\bibnamefont {van
  Kolck}},\ }\bibfield  {title} {\bibinfo {title} {The problem of
  renormalization of chiral nuclear forces},\ }\href
  {https://doi.org/10.3389/fphy.2020.00079} {\bibfield  {journal} {\bibinfo
  {journal} {Frontiers in Physics}\ }\textbf {\bibinfo {volume} {8}},\ \bibinfo
  {pages} {79} (\bibinfo {year} {2020})}\BibitemShut {NoStop}%
\bibitem [{\citenamefont {Lepage}(1997)}]{lepage1997}%
  \BibitemOpen
  \bibfield  {author} {\bibinfo {author} {\bibfnamefont {G.~P.}\ \bibnamefont
  {Lepage}},\ }\bibfield  {title} {\bibinfo {title} {{How to renormalize the
  Schrodinger equation}},\ }in\ \href@noop {} {\emph {\bibinfo {booktitle}
  {{8th Jorge Andre Swieca Summer School on Nuclear Physics}}}}\ (\bibinfo
  {year} {1997})\ \Eprint {https://arxiv.org/abs/nucl-th/9706029}
  {arXiv:nucl-th/9706029} \BibitemShut {NoStop}%
\bibitem [{\citenamefont {Pav\'on~Valderrama}(2011)}]{valderrama2011}%
  \BibitemOpen
  \bibfield  {author} {\bibinfo {author} {\bibfnamefont {M.}~\bibnamefont
  {Pav\'on~Valderrama}},\ }\bibfield  {title} {\bibinfo {title} {Perturbative
  renormalizability of chiral two-pion exchange in nucleon-nucleon
  scattering},\ }\href {https://doi.org/10.1103/PhysRevC.83.024003} {\bibfield
  {journal} {\bibinfo  {journal} {Phys. Rev. C}\ }\textbf {\bibinfo {volume}
  {83}},\ \bibinfo {pages} {024003} (\bibinfo {year} {2011})}\BibitemShut
  {NoStop}%
\bibitem [{\citenamefont {Epelbaum}\ \emph {et~al.}(2015)\citenamefont
  {Epelbaum}, \citenamefont {Gasparyan}, \citenamefont {Gegelia},\ and\
  \citenamefont {Krebs}}]{epelbaum2015b}%
  \BibitemOpen
  \bibfield  {author} {\bibinfo {author} {\bibfnamefont {E.}~\bibnamefont
  {Epelbaum}}, \bibinfo {author} {\bibfnamefont {A.~M.}\ \bibnamefont
  {Gasparyan}}, \bibinfo {author} {\bibfnamefont {J.}~\bibnamefont {Gegelia}},\
  and\ \bibinfo {author} {\bibfnamefont {H.}~\bibnamefont {Krebs}},\ }\bibfield
   {title} {\bibinfo {title} {{$^{1}$S$_{0}$ nucleon-nucleon scattering in the
  modified Weinberg approach}},\ }\href
  {https://doi.org/10.1140/epja/i2015-15071-6} {\bibfield  {journal} {\bibinfo
  {journal} {Eur. Phys. J. A}\ }\textbf {\bibinfo {volume} {51}},\ \bibinfo
  {pages} {71} (\bibinfo {year} {2015})}\BibitemShut {NoStop}%
\bibitem [{\citenamefont {Wiringa}\ \emph {et~al.}(1995)\citenamefont
  {Wiringa}, \citenamefont {Stoks},\ and\ \citenamefont
  {Schiavilla}}]{wiringa1995}%
  \BibitemOpen
  \bibfield  {author} {\bibinfo {author} {\bibfnamefont {R.~B.}\ \bibnamefont
  {Wiringa}}, \bibinfo {author} {\bibfnamefont {V.~G.~J.}\ \bibnamefont
  {Stoks}},\ and\ \bibinfo {author} {\bibfnamefont {R.}~\bibnamefont
  {Schiavilla}},\ }\bibfield  {title} {\bibinfo {title} {Accurate
  nucleon-nucleon potential with charge-independence breaking},\ }\href
  {https://doi.org/10.1103/PhysRevC.51.38} {\bibfield  {journal} {\bibinfo
  {journal} {Phys. Rev. C}\ }\textbf {\bibinfo {volume} {51}},\ \bibinfo
  {pages} {38} (\bibinfo {year} {1995})}\BibitemShut {NoStop}%
\bibitem [{\citenamefont {Machleidt}(2001)}]{machleidt2001}%
  \BibitemOpen
  \bibfield  {author} {\bibinfo {author} {\bibfnamefont {R.}~\bibnamefont
  {Machleidt}},\ }\bibfield  {title} {\bibinfo {title} {{High-precision,
  charge-dependent Bonn nucleon-nucleon potential}},\ }\href
  {https://doi.org/10.1103/PhysRevC.63.024001} {\bibfield  {journal} {\bibinfo
  {journal} {Phys. Rev. C}\ }\textbf {\bibinfo {volume} {63}},\ \bibinfo
  {pages} {024001} (\bibinfo {year} {2001})}\BibitemShut {NoStop}%
\bibitem [{\citenamefont {Pel{\'a}ez}(2016)}]{pelaez2016}%
  \BibitemOpen
  \bibfield  {author} {\bibinfo {author} {\bibfnamefont {J.~R.}\ \bibnamefont
  {Pel{\'a}ez}},\ }\bibfield  {title} {\bibinfo {title} {From controversy to
  precision on the sigma meson: A review on the status of the non-ordinary
  f0(500) resonance},\ }\href {https://doi.org/10.1016/j.physrep.2016.09.001}
  {\bibfield  {journal} {\bibinfo  {journal} {Physics Reports}\ }\textbf
  {\bibinfo {volume} {658}},\ \bibinfo {pages} {1} (\bibinfo {year}
  {2016})}\BibitemShut {NoStop}%
\bibitem [{\citenamefont {Caprini}\ \emph {et~al.}(2006)\citenamefont
  {Caprini}, \citenamefont {Colangelo},\ and\ \citenamefont
  {Leutwyler}}]{Caprini:2005zr}%
  \BibitemOpen
  \bibfield  {author} {\bibinfo {author} {\bibfnamefont {I.}~\bibnamefont
  {Caprini}}, \bibinfo {author} {\bibfnamefont {G.}~\bibnamefont {Colangelo}},\
  and\ \bibinfo {author} {\bibfnamefont {H.}~\bibnamefont {Leutwyler}},\
  }\bibfield  {title} {\bibinfo {title} {{Mass and width of the lowest
  resonance in QCD}},\ }\href {https://doi.org/10.1103/PhysRevLett.96.132001}
  {\bibfield  {journal} {\bibinfo  {journal} {Phys. Rev. Lett.}\ }\textbf
  {\bibinfo {volume} {96}},\ \bibinfo {pages} {132001} (\bibinfo {year}
  {2006})},\ \Eprint {https://arxiv.org/abs/hep-ph/0512364}
  {arXiv:hep-ph/0512364} \BibitemShut {NoStop}%
\bibitem [{\citenamefont {Walecka}(1986)}]{walecka1986}%
  \BibitemOpen
  \bibfield  {author} {\bibinfo {author} {\bibfnamefont {J.~D.}\ \bibnamefont
  {Walecka}},\ }\bibfield  {title} {\bibinfo {title} {The relativistic nuclear
  many-body problem},\ }in\ \href {https://doi.org/10.1007/978-1-4684-5179-5_8}
  {\emph {\bibinfo {booktitle} {New Vistas in Nuclear Dynamics}}},\ \bibinfo
  {series} {NATO ASI Series (Series B: Physics)}, Vol.\ \bibinfo {volume}
  {139},\ \bibinfo {editor} {edited by\ \bibinfo {editor} {\bibfnamefont
  {P.~J.}\ \bibnamefont {Brussaard}}\ and\ \bibinfo {editor} {\bibfnamefont
  {J.~H.}\ \bibnamefont {Koch}}}\ (\bibinfo  {publisher} {Springer},\ \bibinfo
  {address} {Boston, MA},\ \bibinfo {year} {1986})\ pp.\ \bibinfo {pages}
  {229--271}\BibitemShut {NoStop}%
\bibitem [{\citenamefont {Reinhard}(1989)}]{reinhard1989}%
  \BibitemOpen
  \bibfield  {author} {\bibinfo {author} {\bibfnamefont {P.-G.}\ \bibnamefont
  {Reinhard}},\ }\bibfield  {title} {\bibinfo {title} {The relativistic
  mean-field description of nuclei and nuclear dynamics},\ }\href
  {https://doi.org/10.1088/0034-4885/52/4/002} {\bibfield  {journal} {\bibinfo
  {journal} {Reports on Progress in Physics}\ }\textbf {\bibinfo {volume}
  {52}},\ \bibinfo {pages} {439} (\bibinfo {year} {1989})}\BibitemShut
  {NoStop}%
\bibitem [{\citenamefont {Gambhir}\ \emph {et~al.}(1990)\citenamefont
  {Gambhir}, \citenamefont {Ring},\ and\ \citenamefont {Thimet}}]{gambhir1990}%
  \BibitemOpen
  \bibfield  {author} {\bibinfo {author} {\bibfnamefont {Y.~K.}\ \bibnamefont
  {Gambhir}}, \bibinfo {author} {\bibfnamefont {P.}~\bibnamefont {Ring}},\ and\
  \bibinfo {author} {\bibfnamefont {A.}~\bibnamefont {Thimet}},\ }\bibfield
  {title} {\bibinfo {title} {Relativistic mean field theory for finite
  nuclei},\ }\href {https://doi.org/10.1016/0003-4916(90)90330-Q} {\bibfield
  {journal} {\bibinfo  {journal} {Annals of Physics}\ }\textbf {\bibinfo
  {volume} {198}},\ \bibinfo {pages} {132} (\bibinfo {year}
  {1990})}\BibitemShut {NoStop}%
\bibitem [{\citenamefont {Serot}\ and\ \citenamefont
  {Walecka}(1992)}]{serot1992}%
  \BibitemOpen
  \bibfield  {author} {\bibinfo {author} {\bibfnamefont {B.~D.}\ \bibnamefont
  {Serot}}\ and\ \bibinfo {author} {\bibfnamefont {J.~D.}\ \bibnamefont
  {Walecka}},\ }\bibfield  {title} {\bibinfo {title} {Relativistic nuclear
  many-body theory},\ }in\ \href {https://doi.org/10.1007/978-1-4615-3466-2_5}
  {\emph {\bibinfo {booktitle} {Recent Progress in Many-Body Theories}}},\
  \bibinfo {editor} {edited by\ \bibinfo {editor} {\bibfnamefont {T.~L.}\
  \bibnamefont {Ainsworth}}, \bibinfo {editor} {\bibfnamefont {C.~E.}\
  \bibnamefont {Campbell}}, \bibinfo {editor} {\bibfnamefont {B.~E.}\
  \bibnamefont {Clements}},\ and\ \bibinfo {editor} {\bibfnamefont
  {E.}~\bibnamefont {Krotscheck}}}\ (\bibinfo  {publisher} {Springer},\
  \bibinfo {address} {Boston, MA},\ \bibinfo {year} {1992})\ pp.\ \bibinfo
  {pages} {49--92}\BibitemShut {NoStop}%
\bibitem [{\citenamefont {Sugahara}\ and\ \citenamefont
  {Toki}(1994)}]{sugahara1994}%
  \BibitemOpen
  \bibfield  {author} {\bibinfo {author} {\bibfnamefont {Y.}~\bibnamefont
  {Sugahara}}\ and\ \bibinfo {author} {\bibfnamefont {H.}~\bibnamefont
  {Toki}},\ }\bibfield  {title} {\bibinfo {title} {Relativistic mean-field
  theory for unstable nuclei with non-linear $\sigma$ and $\omega$ terms},\
  }\href {https://doi.org/10.1016/0375-9474(94)90923-7} {\bibfield  {journal}
  {\bibinfo  {journal} {Nuclear Physics A}\ }\textbf {\bibinfo {volume}
  {579}},\ \bibinfo {pages} {557} (\bibinfo {year} {1994})}\BibitemShut
  {NoStop}%
\bibitem [{\citenamefont {Donoghue}(2006)}]{donoghue2006}%
  \BibitemOpen
  \bibfield  {author} {\bibinfo {author} {\bibfnamefont {J.~F.}\ \bibnamefont
  {Donoghue}},\ }\bibfield  {title} {\bibinfo {title} {Sigma exchange in the
  nuclear force and effective field theory},\ }\href
  {https://doi.org/10.1016/j.physletb.2006.10.033} {\bibfield  {journal}
  {\bibinfo  {journal} {Physics Letters B}\ }\textbf {\bibinfo {volume}
  {643}},\ \bibinfo {pages} {165} (\bibinfo {year} {2006})}\BibitemShut
  {NoStop}%
\bibitem [{\citenamefont {Crewther}\ and\ \citenamefont
  {Tunstall}(2015)}]{crewther2015}%
  \BibitemOpen
  \bibfield  {author} {\bibinfo {author} {\bibfnamefont {R.~J.}\ \bibnamefont
  {Crewther}}\ and\ \bibinfo {author} {\bibfnamefont {L.~C.}\ \bibnamefont
  {Tunstall}},\ }\bibfield  {title} {\bibinfo {title}
  {{$\mathrm{\ensuremath{\Delta}}I=1/2$ rule for kaon decays derived from QCD
  infrared fixed point}},\ }\href {https://doi.org/10.1103/PhysRevD.91.034016}
  {\bibfield  {journal} {\bibinfo  {journal} {Phys. Rev. D}\ }\textbf {\bibinfo
  {volume} {91}},\ \bibinfo {pages} {034016} (\bibinfo {year}
  {2015})}\BibitemShut {NoStop}%
\bibitem [{\citenamefont {Ma}\ and\ \citenamefont {Rho}(2020)}]{ma2020}%
  \BibitemOpen
  \bibfield  {author} {\bibinfo {author} {\bibfnamefont {Y.-L.}\ \bibnamefont
  {Ma}}\ and\ \bibinfo {author} {\bibfnamefont {M.}~\bibnamefont {Rho}},\
  }\bibfield  {title} {\bibinfo {title} {Towards the hadron–quark continuity
  via a topology change in compact stars},\ }\href
  {https://doi.org/10.1016/j.ppnp.2020.103791} {\bibfield  {journal} {\bibinfo
  {journal} {Progress in Particle and Nuclear Physics}\ }\textbf {\bibinfo
  {volume} {113}},\ \bibinfo {pages} {103791} (\bibinfo {year}
  {2020})}\BibitemShut {NoStop}%
\bibitem [{\citenamefont {Machleidt}\ and\ \citenamefont
  {Entem}(2011)}]{machleidt2011}%
  \BibitemOpen
  \bibfield  {author} {\bibinfo {author} {\bibfnamefont {R.}~\bibnamefont
  {Machleidt}}\ and\ \bibinfo {author} {\bibfnamefont {D.}~\bibnamefont
  {Entem}},\ }\bibfield  {title} {\bibinfo {title} {Chiral effective field
  theory and nuclear forces},\ }\href
  {https://doi.org/10.1016/j.physrep.2011.02.001} {\bibfield  {journal}
  {\bibinfo  {journal} {Physics Reports}\ }\textbf {\bibinfo {volume} {503}},\
  \bibinfo {pages} {1 } (\bibinfo {year} {2011})}\BibitemShut {NoStop}%
\bibitem [{\citenamefont {Ekstr\"om}\ \emph {et~al.}(2013)\citenamefont
  {Ekstr\"om}, \citenamefont {Baardsen}, \citenamefont {Forss\'en},
  \citenamefont {Hagen}, \citenamefont {Hjorth-Jensen}, \citenamefont {Jansen},
  \citenamefont {Machleidt}, \citenamefont {Nazarewicz}, \citenamefont
  {Papenbrock}, \citenamefont {Sarich},\ and\ \citenamefont
  {Wild}}]{ekstrom2013}%
  \BibitemOpen
  \bibfield  {author} {\bibinfo {author} {\bibfnamefont {A.}~\bibnamefont
  {Ekstr\"om}}, \bibinfo {author} {\bibfnamefont {G.}~\bibnamefont {Baardsen}},
  \bibinfo {author} {\bibfnamefont {C.}~\bibnamefont {Forss\'en}}, \bibinfo
  {author} {\bibfnamefont {G.}~\bibnamefont {Hagen}}, \bibinfo {author}
  {\bibfnamefont {M.}~\bibnamefont {Hjorth-Jensen}}, \bibinfo {author}
  {\bibfnamefont {G.~R.}\ \bibnamefont {Jansen}}, \bibinfo {author}
  {\bibfnamefont {R.}~\bibnamefont {Machleidt}}, \bibinfo {author}
  {\bibfnamefont {W.}~\bibnamefont {Nazarewicz}}, \bibinfo {author}
  {\bibfnamefont {T.}~\bibnamefont {Papenbrock}}, \bibinfo {author}
  {\bibfnamefont {J.}~\bibnamefont {Sarich}},\ and\ \bibinfo {author}
  {\bibfnamefont {S.~M.}\ \bibnamefont {Wild}},\ }\bibfield  {title} {\bibinfo
  {title} {Optimized chiral nucleon-nucleon interaction at
  next-to-next-to-leading order},\ }\href
  {https://doi.org/10.1103/PhysRevLett.110.192502} {\bibfield  {journal}
  {\bibinfo  {journal} {Phys. Rev. Lett.}\ }\textbf {\bibinfo {volume} {110}},\
  \bibinfo {pages} {192502} (\bibinfo {year} {2013})}\BibitemShut {NoStop}%
\bibitem [{\citenamefont {Ekstr\"om}\ \emph {et~al.}(2015)\citenamefont
  {Ekstr\"om}, \citenamefont {Jansen}, \citenamefont {Wendt}, \citenamefont
  {Hagen}, \citenamefont {Papenbrock}, \citenamefont {Carlsson}, \citenamefont
  {Forss\'en}, \citenamefont {Hjorth-Jensen}, \citenamefont {Navr\'atil},\ and\
  \citenamefont {Nazarewicz}}]{ekstrom2015a}%
  \BibitemOpen
  \bibfield  {author} {\bibinfo {author} {\bibfnamefont {A.}~\bibnamefont
  {Ekstr\"om}}, \bibinfo {author} {\bibfnamefont {G.~R.}\ \bibnamefont
  {Jansen}}, \bibinfo {author} {\bibfnamefont {K.~A.}\ \bibnamefont {Wendt}},
  \bibinfo {author} {\bibfnamefont {G.}~\bibnamefont {Hagen}}, \bibinfo
  {author} {\bibfnamefont {T.}~\bibnamefont {Papenbrock}}, \bibinfo {author}
  {\bibfnamefont {B.~D.}\ \bibnamefont {Carlsson}}, \bibinfo {author}
  {\bibfnamefont {C.}~\bibnamefont {Forss\'en}}, \bibinfo {author}
  {\bibfnamefont {M.}~\bibnamefont {Hjorth-Jensen}}, \bibinfo {author}
  {\bibfnamefont {P.}~\bibnamefont {Navr\'atil}},\ and\ \bibinfo {author}
  {\bibfnamefont {W.}~\bibnamefont {Nazarewicz}},\ }\bibfield  {title}
  {\bibinfo {title} {Accurate nuclear radii and binding energies from a chiral
  interaction},\ }\href {https://doi.org/10.1103/PhysRevC.91.051301} {\bibfield
   {journal} {\bibinfo  {journal} {Phys. Rev. C}\ }\textbf {\bibinfo {volume}
  {91}},\ \bibinfo {pages} {051301} (\bibinfo {year} {2015})}\BibitemShut
  {NoStop}%
\bibitem [{\citenamefont {Entem}\ \emph {et~al.}(2015)\citenamefont {Entem},
  \citenamefont {Kaiser}, \citenamefont {Machleidt},\ and\ \citenamefont
  {Nosyk}}]{entem2015}%
  \BibitemOpen
  \bibfield  {author} {\bibinfo {author} {\bibfnamefont {D.~R.}\ \bibnamefont
  {Entem}}, \bibinfo {author} {\bibfnamefont {N.}~\bibnamefont {Kaiser}},
  \bibinfo {author} {\bibfnamefont {R.}~\bibnamefont {Machleidt}},\ and\
  \bibinfo {author} {\bibfnamefont {Y.}~\bibnamefont {Nosyk}},\ }\bibfield
  {title} {\bibinfo {title} {Peripheral nucleon-nucleon scattering at fifth
  order of chiral perturbation theory},\ }\href
  {https://doi.org/10.1103/PhysRevC.91.014002} {\bibfield  {journal} {\bibinfo
  {journal} {Phys. Rev. C}\ }\textbf {\bibinfo {volume} {91}},\ \bibinfo
  {pages} {014002} (\bibinfo {year} {2015})}\BibitemShut {NoStop}%
\bibitem [{\citenamefont {Jiang}\ \emph {et~al.}(2020)\citenamefont {Jiang},
  \citenamefont {Ekstr\"om}, \citenamefont {Forss\'en}, \citenamefont {Hagen},
  \citenamefont {Jansen},\ and\ \citenamefont {Papenbrock}}]{jiang2020}%
  \BibitemOpen
  \bibfield  {author} {\bibinfo {author} {\bibfnamefont {W.~G.}\ \bibnamefont
  {Jiang}}, \bibinfo {author} {\bibfnamefont {A.}~\bibnamefont {Ekstr\"om}},
  \bibinfo {author} {\bibfnamefont {C.}~\bibnamefont {Forss\'en}}, \bibinfo
  {author} {\bibfnamefont {G.}~\bibnamefont {Hagen}}, \bibinfo {author}
  {\bibfnamefont {G.~R.}\ \bibnamefont {Jansen}},\ and\ \bibinfo {author}
  {\bibfnamefont {T.}~\bibnamefont {Papenbrock}},\ }\bibfield  {title}
  {\bibinfo {title} {Accurate bulk properties of nuclei from $a=2$ to
  $\ensuremath{\infty}$ from potentials with $\mathrm{\ensuremath{\Delta}}$
  isobars},\ }\href {https://doi.org/10.1103/PhysRevC.102.054301} {\bibfield
  {journal} {\bibinfo  {journal} {Phys. Rev. C}\ }\textbf {\bibinfo {volume}
  {102}},\ \bibinfo {pages} {054301} (\bibinfo {year} {2020})}\BibitemShut
  {NoStop}%
\bibitem [{\citenamefont {Hoferichter}\ \emph {et~al.}(2015)\citenamefont
  {Hoferichter}, \citenamefont {Ruiz~de Elvira}, \citenamefont {Kubis},\ and\
  \citenamefont {Mei\ss{}ner}}]{hoferichter2015}%
  \BibitemOpen
  \bibfield  {author} {\bibinfo {author} {\bibfnamefont {M.}~\bibnamefont
  {Hoferichter}}, \bibinfo {author} {\bibfnamefont {J.}~\bibnamefont {Ruiz~de
  Elvira}}, \bibinfo {author} {\bibfnamefont {B.}~\bibnamefont {Kubis}},\ and\
  \bibinfo {author} {\bibfnamefont {U.-G.}\ \bibnamefont {Mei\ss{}ner}},\
  }\bibfield  {title} {\bibinfo {title} {Matching pion-nucleon roy-steiner
  equations to chiral perturbation theory},\ }\href
  {https://doi.org/10.1103/PhysRevLett.115.192301} {\bibfield  {journal}
  {\bibinfo  {journal} {Phys. Rev. Lett.}\ }\textbf {\bibinfo {volume} {115}},\
  \bibinfo {pages} {192301} (\bibinfo {year} {2015})}\BibitemShut {NoStop}%
\bibitem [{\citenamefont {Hoferichter}\ \emph {et~al.}(2016)\citenamefont
  {Hoferichter}, \citenamefont {{Ruiz de Elvira}}, \citenamefont {Kubis},\ and\
  \citenamefont {Mei{\ss}ner}}]{hoferichter2016}%
  \BibitemOpen
  \bibfield  {author} {\bibinfo {author} {\bibfnamefont {M.}~\bibnamefont
  {Hoferichter}}, \bibinfo {author} {\bibfnamefont {J.}~\bibnamefont {{Ruiz de
  Elvira}}}, \bibinfo {author} {\bibfnamefont {B.}~\bibnamefont {Kubis}},\ and\
  \bibinfo {author} {\bibfnamefont {U.-G.}\ \bibnamefont {Mei{\ss}ner}},\
  }\bibfield  {title} {\bibinfo {title} {Roy-steiner-equation analysis of
  pion-nucleon scattering},\ }\href
  {https://doi.org/10.1016/j.physrep.2016.02.002} {\bibfield  {journal}
  {\bibinfo  {journal} {Phys. Rept.}\ }\textbf {\bibinfo {volume} {625}},\
  \bibinfo {pages} {1 } (\bibinfo {year} {2016})}\BibitemShut {NoStop}%
\bibitem [{\citenamefont {Epelbaum}\ \emph
  {et~al.}(2004{\natexlab{a}})\citenamefont {Epelbaum}, \citenamefont
  {Gloeckle},\ and\ \citenamefont {Meissner}}]{Epelbaum:2003gr}%
  \BibitemOpen
  \bibfield  {author} {\bibinfo {author} {\bibfnamefont {E.}~\bibnamefont
  {Epelbaum}}, \bibinfo {author} {\bibfnamefont {W.}~\bibnamefont {Gloeckle}},\
  and\ \bibinfo {author} {\bibfnamefont {U.-G.}\ \bibnamefont {Meissner}},\
  }\bibfield  {title} {\bibinfo {title} {{Improving the convergence of the
  chiral expansion for nuclear forces. 1. Peripheral phases}},\ }\href
  {https://doi.org/10.1140/epja/i2003-10096-0} {\bibfield  {journal} {\bibinfo
  {journal} {Eur. Phys. J. A}\ }\textbf {\bibinfo {volume} {19}},\ \bibinfo
  {pages} {125} (\bibinfo {year} {2004}{\natexlab{a}})},\ \Eprint
  {https://arxiv.org/abs/nucl-th/0304037} {arXiv:nucl-th/0304037} \BibitemShut
  {NoStop}%
\bibitem [{\citenamefont {Epelbaum}\ \emph
  {et~al.}(2004{\natexlab{b}})\citenamefont {Epelbaum}, \citenamefont
  {Gl\"ockle},\ and\ \citenamefont {Mei\ss{}ner}}]{epelbaum04}%
  \BibitemOpen
  \bibfield  {author} {\bibinfo {author} {\bibfnamefont {E.}~\bibnamefont
  {Epelbaum}}, \bibinfo {author} {\bibfnamefont {W.}~\bibnamefont
  {Gl\"ockle}},\ and\ \bibinfo {author} {\bibfnamefont {U.-G.}\ \bibnamefont
  {Mei\ss{}ner}},\ }\bibfield  {title} {\bibinfo {title} {Improving the
  convergence of the chiral expansion for nuclear forces - ii: Low phases and
  the deuteron},\ }\href {https://doi.org/10.1140/epja/i2003-10129-8}
  {\bibfield  {journal} {\bibinfo  {journal} {Eur. Phys. J. A}\ }\textbf
  {\bibinfo {volume} {19}},\ \bibinfo {pages} {401} (\bibinfo {year}
  {2004}{\natexlab{b}})}\BibitemShut {NoStop}%
\bibitem [{\citenamefont {P\'erez}\ \emph {et~al.}(2013)\citenamefont
  {P\'erez}, \citenamefont {Amaro},\ and\ \citenamefont
  {Arriola}}]{navarro2013}%
  \BibitemOpen
  \bibfield  {author} {\bibinfo {author} {\bibfnamefont {R.~N.}\ \bibnamefont
  {P\'erez}}, \bibinfo {author} {\bibfnamefont {J.~E.}\ \bibnamefont {Amaro}},\
  and\ \bibinfo {author} {\bibfnamefont {E.~R.}\ \bibnamefont {Arriola}},\
  }\bibfield  {title} {\bibinfo {title} {Coarse-grained potential analysis of
  neutron-proton and proton-proton scattering below the pion production
  threshold},\ }\href {https://doi.org/10.1103/PhysRevC.88.064002} {\bibfield
  {journal} {\bibinfo  {journal} {Phys. Rev. C}\ }\textbf {\bibinfo {volume}
  {88}},\ \bibinfo {pages} {064002} (\bibinfo {year} {2013})}\BibitemShut
  {NoStop}%
\bibitem [{\citenamefont {Kaiser}\ \emph {et~al.}(1998)\citenamefont {Kaiser},
  \citenamefont {Gerstend{\"o}rfer},\ and\ \citenamefont {Weise}}]{kaiser1998}%
  \BibitemOpen
  \bibfield  {author} {\bibinfo {author} {\bibfnamefont {N.}~\bibnamefont
  {Kaiser}}, \bibinfo {author} {\bibfnamefont {S.}~\bibnamefont
  {Gerstend{\"o}rfer}},\ and\ \bibinfo {author} {\bibfnamefont
  {W.}~\bibnamefont {Weise}},\ }\bibfield  {title} {\bibinfo {title}
  {Peripheral nn-scattering: role of delta-excitation, correlated two-pion and
  vector meson exchange},\ }\href
  {https://doi.org/10.1016/S0375-9474(98)00234-6} {\bibfield  {journal}
  {\bibinfo  {journal} {Nucl. Phys. A}\ }\textbf {\bibinfo {volume} {637}},\
  \bibinfo {pages} {395 } (\bibinfo {year} {1998})}\BibitemShut {NoStop}%
\bibitem [{\citenamefont {Krebs}\ \emph {et~al.}(2007)\citenamefont {Krebs},
  \citenamefont {Epelbaum},\ and\ \citenamefont {Mei{\ss}ner}}]{krebs2007}%
  \BibitemOpen
  \bibfield  {author} {\bibinfo {author} {\bibfnamefont {H.}~\bibnamefont
  {Krebs}}, \bibinfo {author} {\bibfnamefont {E.}~\bibnamefont {Epelbaum}},\
  and\ \bibinfo {author} {\bibfnamefont {U.~G.}\ \bibnamefont {Mei{\ss}ner}},\
  }\bibfield  {title} {\bibinfo {title} {Nuclear forces with
  $\mathrm{\ensuremath{\Delta}}$ excitations up to next-to-next-to-leading
  order, part i: Peripheral nucleon-nucleon waves},\ }\href
  {https://doi.org/10.1140/epja/i2007-10372-y} {\bibfield  {journal} {\bibinfo
  {journal} {Eur. Phys. J. A}\ }\textbf {\bibinfo {volume} {32}},\ \bibinfo
  {pages} {127} (\bibinfo {year} {2007})}\BibitemShut {NoStop}%
\end{thebibliography}
\end{document}